\documentclass[aps,prd,reprint,superscriptaddress,longbibliography,showpacs]{revtex4-2}

\usepackage{graphicx}
\usepackage{dcolumn}
\usepackage{bm}

\usepackage[figurename=Figure]{caption}
\usepackage{float}    
\usepackage{verbatim} 
\usepackage{amsmath}  
\usepackage{upgreek} 
\usepackage{amssymb}  
\usepackage{listings} 
\usepackage{graphics}
\usepackage{enumitem} 
\usepackage{latexsym,epsfig}
\usepackage{subcaption}
\usepackage{physics}
\usepackage{stackrel}

\usepackage{placeins} 

\usepackage[colorlinks=true,breaklinks=true,allcolors=blue]{hyperref}
\usepackage{lipsum}
\usepackage{dsfont}

\usepackage[scr=boondox]{mathalfa}

\def\Oo{\ensuremath{{\cal O}}} 
\def\Jj{\ensuremath{{\cal J}}} 

\def\Ww{\ensuremath{{\cal W}}}

\def\Hh{\ensuremath{{\cal H}}} 
\def\Nn{\ensuremath{{N}}}

\def\Ww{\ensuremath{{\cal W}}}

\def\Cc{\ensuremath{{\cal C}}}
\def\Dd{\ensuremath{{\cal D}}}
\def\Ff{\ensuremath{{\cal F}}}
\def\Kk{\ensuremath{{\cal K}}}
\def\Tt{\ensuremath{{\cal T}}} 
\def\Gg{\ensuremath{{\cal G}}} 
\def\Zz{\ensuremath{{\cal Z}}}

\def\Qq{\ensuremath{{\cal Q}}}

\usepackage[scr=boondox]{mathalfa}

\def\p{\ensuremath{{\partial}}}
\newcommand{\sgn}{\text{sgn}}
\def\i{\ensuremath{{\imath}}}

\def\Im{\ensuremath{{\operatorname{Im}}}}
\def\Re{\ensuremath{{\operatorname{Re}}}}

\newcommand\ex[1]{ \langle #1 \rangle }

\def\l{{\color{green}\lambda}}

\def\ll{{\color{green}\boldsymbol{\lambda}}}
\def\et{{\color{red}\boldsymbol{\eta}}}
\def\llbar{\bar{{\color{green}\boldsymbol{\lambda}}}}
\def\etbar{\bar{{\color{red}\boldsymbol{\eta}}}}

\newcommand{\be}{\begin{equation}}
	\newcommand{\ee}{\end{equation}}
\newcommand{\bea}{\begin{equation}\begin{aligned}}
		\newcommand{\eea}{\end{aligned}\end{equation}}
\newcommand{\ben}{\begin{enumerate}}
	\newcommand{\een}{\end{enumerate}}

\DeclareDocumentCommand{\nint}{ O{} O{} m }{\ensuremath{ \int_{\mbox{\scriptsize $#1$}}^{\mbox{\scriptsize$#2$}}\!\!\! \mbox{\small $\,\mathrm{d}#3$\! }}}

\usepackage{tikz,bm} 
\usepackage{circuitikz}
\usetikzlibrary{decorations.markings}
\usetikzlibrary{decorations.pathreplacing,calligraphy}

\definecolor{mycolor}{rgb}{1,0.2,0.3}
\definecolor{brightgreen}{rgb}{0.4, 1.0, 0.0}
\definecolor{britishracinggreen}{rgb}{0.0, 0.26, 0.15}
\definecolor{cadmiumgreen}{rgb}{0.0, 0.42, 0.24}
\definecolor{ceruleanblue}{rgb}{0.16, 0.32, 0.75}
\definecolor{darkelectricblue}{rgb}{0.33, 0.41, 0.47}
\definecolor{darkpowderblue}{rgb}{0.0, 0.2, 0.6}
\definecolor{darktangerine}{rgb}{1.0, 0.66, 0.07}
\definecolor{emerald}{rgb}{0.31, 0.78, 0.47}
\definecolor{palatinatepurple}{rgb}{0.41, 0.16, 0.38}
\definecolor{pastelviolet}{rgb}{0.8, 0.6, 0.79}

\newcommand*\circled[1]{\tikz[baseline=(char.base)]{
            \node[shape=circle,draw,inner sep=2pt] (char) {#1};}}
\usepackage{orcidlink}
\newcommand{\orcidLouw}{\orcidlink{0000-0002-5111-840X}}
\begin{document}
	
	\preprint{APS/123-QED}
	
	\title{Current Correlations and Conductivity in SYK-Like Systems: An Analytical Study}
	\author{Rishabh Jha}
	\email{rishabh.jha@uni-goettingen.de}
	\affiliation{%
		Institute for Theoretical Physics, Georg-August-Universit\"{a}t G\"{o}ttingen, Friedrich-Hund-Platz 1, 37077 G\"ottingen, Germany
	}%
	
\author{Stefan Kehrein}
\affiliation{%
		Institute for Theoretical Physics, Georg-August-Universit\"{a}t G\"{o}ttingen, Friedrich-Hund-Platz 1, 37077 G\"ottingen, Germany
	}%
\author{Jan C. Louw\orcidLouw}
	\affiliation{%
		Institute for Theoretical Physics, Georg-August-Universit\"{a}t G\"{o}ttingen, Friedrich-Hund-Platz 1, 37077 G\"ottingen, Germany
	}%
    \affiliation{Max Planck Institute of Quantum Optics, 85748 Garching, Germany}

	\begin{abstract}
We present a functional-based approach to compute thermal expectation values for actions expressed in the $G-\Sigma$ formalism, applicable to any time sequence ordering. Utilizing this framework, we analyze the linear response to an electric field in various Sachdev-Ye-Kitaev (SYK) chains. We consider the SYK chain where each dot is a complex $q/2$-body interacting SYK model, and we allow for $r/2$-body nearest-neighbor hopping where $r=\kappa q$. We find exact analytical expressions in the large-$q$ limit for conductivities across all temperatures at leading order in $1/q$ for three cases, namely $\kappa = \{ 1/2, 1, 2\}$. When $\kappa = \{1/2, 1\}$, we observe linear-in-temperature $T$ resistivities at low temperatures, indicative of strange metal behavior. Conversely, when $\kappa = 2$, the resistivity diverges as a power law at low temperatures, namely as $1/T^2$, resembling insulating behavior. As $T$ increases, there is a crossover to Fermi liquid behavior ($\sim T^2$) at the minimum resistivity. Beyond this, another crossover occurs to strange metal behavior ($\sim T$). In comparison to previous linear-in-$T$ results in the literature, we also show that the resistivity behavior exists even below the MIR bound, indicating a true strange metal instead of a bad metal. In particular, we find for the $\kappa = 2$ case a smooth crossover from an insulating phase to a Fermi liquid behavior to a true strange metal and eventually becoming a bad metal as temperature increases. We extend and generalize previously known results on resistivities to all temperatures, do a comparative analysis across the three models where we highlight the universal features and invoke scaling arguments to create a physical picture out of our analyses. Remarkably, we find a universal maximum DC conductivity across all three models when the hopping coupling strength becomes large.


\end{abstract}
	
	\maketitle
	

	\section{Introduction and results}
 \label{introduction section}

 The exploration of quantum systems out of equilibrium continues to unveil significant insights into the fundamental mechanisms of thermalization, transport, and quantum many-body chaos within condensed matter systems. The Sachdev-Ye-Kitaev (SYK) model, describing fermions with random all-to-all interactions, has become a focal point in the study of quantum chaos and non-Fermi liquid behavior. One of the interesting features of SYK models is their approximate conformal field theory (CFT) behavior at low temperatures, coupled with maximal many-body chaos. This model's connection to holography, particularly its duality to extremal charged black holes with AdS$_2$ horizons, underscores its significance in the theoretical exploration of strange metals and quantum criticality \cite{Sachdev2015,Maldacena2016Dec,Kitaev2015,Rosenhaus2019,Louw2023Dec}.

The strange metal phase associated with high-$T_c$ superconductivity is of particular interest, yet solvable models capturing this phase are scarce. The complex SYK variant, which includes a conserved global U(1) charge, allows one to also consider charge transport by coupling the dots into a chain or some higher dimensional lattice \cite{Song2017Nov,Chen2017Jul,Khveshchenko2018Jul,Davison2017,Gu2017May,Burdin2002Jul,Parcollet1999Feb,Zhang2017Nov,Cai2018Jan,Parcollet1999Feb,Burdin2002Jul,Khveshchenko2018Jul,Jian2017Sep,Zanoci2022Jun,Chowdhury2022Sep,Gu2020Feb, Zhang2021Nov,Jha2023}.

Particularly with $q$-body interactions and taking $q$ to be large, this SYK chain offers an analytical solution for studying non-equilibrium physics \cite{Louw2022,Eberlein2017Nov} and transport \cite{Gu2020Feb, Davison2017, Zhang2021Nov,Zanoci2022Jun,Jha2023}. Here, we extend this analysis to include transport of order $\kappa q$, presenting solutions for $\kappa =\{ 1/2, 1, 2\} $. While the $\kappa=1$ case was solved and analyzed in \cite{Zanoci2022Jun}, we provide two novel solutions, namely $\kappa =1/2,2$. These solvable chains displaying Planckian dynamics provide a fertile ground for addressing holography questions, such as the correspondence between transport along the SYK dots and its gravitational dual. Again, by considering $q$ to be large, we find exact analytical solutions for any temperature.

In this work, we employ a functional-based approach to compute thermal expectation values for $G-\Sigma$ actions, facilitating the study of linear responses to electric fields in SYK chains. This methodology allows us to derive and compare conductivities across all temperatures. For hopping orders that are half or equal to the on-site interaction orders, we find linear-in-temperature $T$ resistivities at low temperatures, indicative of strange metal behavior. When the hopping order exceeds the on-site order by two, the resistivity diverges as a power law at low temperatures, $\rho_{\text{DC}} \sim T^{-2}$, resembling insulating behavior. Across all cases, a universal maximum DC conductivity, $\sigma_{\text{DC}}^{\text{max}} = N \pi /8$, is observed, where $N$ is the number of fermion charge carriers or flavors per quantum dot. Additionally, we discuss the Mott-Ioffe-Regel (MIR) limit, $\rho_{\text{MIR}} = \frac{2\pi \hbar}{q_e^2 N a}$, which is $2\pi/N$ in our units \cite{Werman2016Feb,Hartnoll2015Jan}. This typical upper limit on resistivity is related to the scattering rate and mean free path such that the mean-free path of charge carriers becomes equal to that of the lattice spacing \cite{Hartnoll2015Jan}. For so-called ``bad'' metals, this bound can be over-shot, often linearly in $T$. Such a linear-in-$T$ resistivity that is also below the MIR bound is what defines a ``strange'' metal instead of just a bad metal which is in line with Refs. \cite{Chowdhury2022Sep, YSYK-strange-bad-metal}. Our minimum resistivity of $8/( N \pi )$ is less than $2\pi/N$. Thus, our results show linear-in-$T$ both below and above the MIR bound, indicating a strange metal as well as a bad metal (see Fig. \ref{fig:Res}). In this sense, it yields novel results in comparison to the bad metal SYK chain considered in \cite{Song2017Nov}. In fact, we find the $\kappa =2$ case particularly enlightening in light of difficulties \cite{Chowdhury2022Sep, sachdev-essay} in finding smooth crossovers from a Fermi liquid description to a true strange metal behavior that eventually becomes bad metal at even higher temperatures. 
We find such smooth crossovers analytically in the thermodynamic limit for the $\kappa = 2$ case starting from an insulating phase where $\rho \sim T^{-2}$ at low temperatures before the onset of the Fermi liquid behavior, therefore enabling further investigation in this direction as discussed in the outlook. Finding such an electrical insulating phase at low temperatures is not a complete surprise given that the uncharged Majorana versions of the chain has a thermal insulating phase reflected is its diffusion constant \cite{Jian2017Sep}. However they only study the thermal transport properties of the Majorana model for the $\kappa = 2$ case while we study the electrical transport properties for the more general complex fermions. 

In the context of thermodynamic phase diagrams, high-temperature behavior emerges when effective couplings, exponentially suppressed by the charge density, become small \cite{Louw2023Feb,Louw2023Dec}. This occurs at large charge densities, corresponding to large chemical potentials, which can also be interpreted as a uniform electric field. This is what we consider in this work, where analyze the linear response by applying an electric field to the uniformly coupled SYK chain at equilibrium and calculate the conductivities. 

Using the scaling arguments for interacting fermions \cite{Shankar1994Jan}, we realize that as temperature tends to zero, the term in the Hamiltonian containing less number of fermions dominates the overall behavior. In our case, this argument would amount to the situation that, for $\kappa = 1/2$, the hopping term dominates the on-site large-$q$ SYK term and we observe a conducting phase. For $\kappa =2$ case, this would result in the on-site interaction dominating over the hopping term, thereby leading to an insulating behavior. Only for $\kappa = 1$ case, we would have a competition between the on-site term and the hopping term, where no \textit{a priori} conclusion can be drawn about the nature of resistivity. Having said that, the scaling argument is not rigorously established away from critical point for SYK-like systems which are models without quasiparticles and therefore, it cannot be taken as a substitute for mathematical calculations to establish the nature of transport. We show via detailed mathematical calculations that despite there being no proof for scaling arguments, the picture holds, and we do find conducting behavior for the $\kappa = 1/2$ case and insulating behavior for the $\kappa = 2$ case at low temperatures. In addition, we find a conducting phase at low temperatures for the $\kappa =1$ case for which we cannot get insights from the scaling arguments. This is made precise in Section \ref{a note on scaling arguments}.

The structure of the paper is as follows. Section \ref{model section} introduces the three SYK chains we consider, along with the mapping of a uniformly coupled chain at (Eq. \eqref{model equation}) to a single SYK dot (Eq. \eqref{model equivalent equation}). Section \ref{analytical solution subsection} provides exact analytical results to leading order in $1/q$ for the three cases of interest, where we also discuss the scaling arguments in detail. Section \ref{keldysh contour deformation section} introduces the Keldysh contour deformations and derives general relations for contour deformations corresponding to linearly perturbing the system at equilibrium to study the linear response. We deform both the forward and the backward Keldysh contour and derive general relations that can be applied to SYK-like systems where action is written in terms of $G-\Sigma$ formalism. In Section \ref{current section}, we define the current operator using the continuity equation and apply an electric field to study the linear response. Section \ref{current correlations section} evaluates the current-current correlations, and Section \ref{conductivity section} evaluates the DC conductivities and resistivities across all temperatures, where we specialize to low-temperature regime to study the strange metal behavior. We also compare and contrast the three models against the MIR bound and prove that the linear-in-$T$ resistivity indeed signals a true strange metal behavior for some temperature regimes (in contrast to the bad metal behavior); see Fig. \ref{fig:Res}. We conclude in Section \ref{conclusion section} with an outlook on further research lines using our Keldysh functional-based approach for SYK-like systems.

 \section{Model}
 \label{model section}

We consider a uniformly coupled chain of $L$ SYK dots
\begin{equation}
    \hat{\Hh} = \sum_{i=1}^L \left( \hat{\Hh}_i + \hat{\Hh}_{i\to i+1}+ \hat{\Hh}_{i\to i+1}^\dag\right)
    \label{model equation}
\end{equation}
where the on-site $q$-body interaction is given by
\begin{equation}
    \hat{\Hh}_{i}= J \hspace{-1mm} \sum\limits_{\substack{ \{\bm{\mu}\}_1^{q/2} \\ \{\bm{\nu}\}_1^{q/2} }} \hspace{-2mm}X(i)^{\bm{\mu}}_{\bm{\nu}} c^{\dag}_{i;\mu_1} \cdots c^{\dag}_{i; \mu_{q/2}} c_{i;\nu_{q/2}}^{\vphantom{\dag}} \cdots c_{i;\nu_1}^{\vphantom{\dag}}
\end{equation}
summing over $\{\bm{\mu}\}_{1}^{q/2} \equiv 1\le \mu_1<\cdots< \mu_{q/2}\le\Nn$ and $\{\bm{\nu}\}_{1}^{q/2} \equiv 1\le \nu_1<\cdots< \nu_{q/2}\le\Nn$. Transport between nearest neighbors is mediated by $r = \kappa q$-body SYK-like Hamiltonian given by
\begin{equation}
\hat{\Hh}_{i\to i+1}\equiv D \sum\limits_{\substack{ \{\bm{\mu}\}_1^{r/2} \\ \{\bm{\nu}\}_1^{r/2} }} \hspace{-2mm}Y(i)^{\bm{\mu}}_{\bm{\nu}} c^{\dag}_{i+1;\mu_1} \cdots c^{\dag}_{i+1; \mu_{\frac{r}{2}}} c_{i;\nu_{\frac{r}{2}}}^{\vphantom{\dag}} \cdots c_{i;\nu_1}^{\vphantom{\dag}}.
\label{hi to i+1}
	\end{equation}
Latin indices denote the lattice sites in real space while Greek indices denote the flavors. We are interested in the large-$N$ limit followed by large-$q$ limit where our analysis will be at leading order in $1/q$. 

Both $X(i)^{\bm{\mu}}_{\bm{\nu}}$ and $Y(i)^{\bm{\mu}}_{\bm{\nu}}$ are random variables derived from Gaussian ensembles with zero mean and the following variance:
\begin{equation}
\begin{aligned}
& \overline{|X|^2}=\frac{q^{-2}((q / 2)!)^2}{(N / 2)^{q-1}} \\
& \overline{|Y|^2}=\frac{1}{q} \frac{(1 / r) ((r / 2)!)^2}{(N / 2)^{r-1}} .
\end{aligned}
\end{equation}
This system has a conserved total charge per flavor
 \begin{equation}
     \hat{\Qq} = \sum_{j=1}^L \hat{\Qq}_{j}, 
     \label{total charge}
 \end{equation}
where the local charge density is defined by
 \begin{equation}
\hat{\mathcal{Q}}_j \equiv \frac{1}{N} \sum_{\alpha=1}^N\left[c_{j ; \alpha}^{\dagger} c_{j ; \alpha}-1 / 2\right]
\label{local charge density definition}
\end{equation}
and is associated with a total extensive (in $L$ and $N$) current operator
     \begin{equation}
        \hat{I} =  \frac{\i r \hat{\Hh}_{\rightarrow}  - \i r  \hat{\Hh}_{\rightarrow}^\dag}{2} \label{currentOp}
    \end{equation}
 defined by the continuity equation as derived later in the manuscript in Section \ref{continuity equation subsection}. Here we have defined
 \begin{equation}
     \hat{\Hh}_{\rightarrow} \equiv \sum\limits_{i=1}^L \Hh_{i \to i+1}
     \label{H rightarrow defined}
 \end{equation}
 and together with the transport terms in Eq. \eqref{model equation}, we can define
 \begin{equation}
     \hat{\Hh}_{\text{dot}} \equiv \sum\limits_{i=1}^L \hat{\Hh}_i, \quad \hat{\Hh}_{\text{trans}} \equiv \hat{\Hh}_{\rightarrow} + \hat{\Hh}_{\rightarrow}^\dagger
     \label{H dot and H trans defined}
 \end{equation}
 thereby re-writing the model in Eq. \eqref{model equation} as
 \begin{equation}
     \hat{\Hh} = \hat{\Hh}_{\text{dot}} + \hat{\Hh}_{\text{trans}}
 \end{equation}

 We are interested in extracting out the thermal expectation values of $\ex{\hat{\Hh}_{\rightarrow}(\tau)}$ where the expectation value is taken with respect to the thermal state $\rho \propto e^{-\beta [\hat{\Hh}-N\mu \hat{\Qq}]}$. In general, we will denote such expectation values by the operator symbol without the hat, e.g., $\Qq = \ex{\hat{\Qq}}$.

As has been shown in Appendix A of Ref. \cite{Louw2023Dec}, the uniformly coupled Hamiltonian at equilibrium in Eq. \eqref{model equation} is equivalent to a single-dot SYK whose Hamiltonian is given by
    \begin{equation}
		\hat{\Hh} =  J_q\hat{\Hh}_q + K_{\kappa q} \hat{\Hh}_{ \kappa q}
		\label{model equivalent equation}
	\end{equation}
where we identify $J^2 \to J_q^2$ and $2 |D|^2 \to K_{\kappa q}^2$. Here $\Hh_{\kappa q}$ is large-$\kappa q$ complex SYK model Hamiltonian given by
	\begin{equation}
		\hat{\Hh}_{\kappa q}= \hspace{-1mm} \sum\limits_{\substack{ \{\bm{\mu}\}_1^{\kappa q/2} \\ \{\bm{\nu}\}_1^{\kappa q/2} }}^N \hspace{-2mm} Z^{\bm{\mu}}_{\bm{\nu}} c^{\dag}_{\mu_1} \cdots c^{\dag}_{\mu_{\kappa q/2}} c_{\nu_{\kappa q/2}}^{\vphantom{\dag}} \cdots c_{\nu_1}^{\vphantom{\dag}} 
	\end{equation}
	summing over $\{\bm{\mu}\}_{1}^{\kappa q/2} \equiv 1\le \mu_1<\cdots< \mu_{\kappa q/2}\le\Nn$ and $\{\bm{\nu}\}_{1}^{\kappa q/2} \equiv 1\le \nu_1<\cdots< \nu_{\kappa q/2}\le\Nn$. Here $Z^{\bm{\mu}}_{\bm{\nu}}$ is a random matrix whose components are derived from a Gaussian ensemble with zero mean and variance given by
	\begin{equation}
		\overline{|Z|^2} =  \frac{(\kappa q)^{-2}((\kappa q / 2) !)^2}{(N / 2)^{\kappa q-1}} 
		\label{variance of Z random matrix}
	\end{equation}
The total charge in Eq. \eqref{total charge} is associated with the total charge of the dot using this mapping. We are assuming coupling constants to be real throughout.

\section{Exact analytical solution at leading order in 1/q}
\label{analytical solution subsection}

We are interested in calculating the current-current correlations and conductivity of uniformly coupled chain in Eq. \eqref{model equation} where we consider the equilibrium situation and will apply the linear response theory \cite{Snoke2020Jan}. We will be interested in three cases, namely $\kappa = \{ \frac{1}{2}, 1, 2\}$ where we solve the chain using the mapping to a dot as in Eq. \eqref{model equivalent equation} and use the exact analytical solution at leading order in $1/q$ to calculate the aforementioned quantities. We will be doing calculations in real time where we will employ the machinery of Schwinger-Keldysh formalism \cite{Stefanucci2013Mar}. 

We know that our uniformly coupled chain at equilibrium in Eq. \eqref{model equation} can be mapped onto a dot as in Eq. \eqref{model equivalent equation} with the identification $J^2 \to J_q^2$ and $2 |D|^2 \to J_{\kappa q}^2$, therefore, we focus on the latter as a proxy to analyze the chain. We start by calculating the Schwinger-Dyson equations for the system in Eq. \eqref{model equivalent equation} where we follow the standard large-$N$ calculations followed by large-$q$ limit \cite{Maldacena2016Nov} to calculate the effective action by integrating out the fermions that leaves us with the following expression for the partition function:
\begin{equation}
\mathcal{Z}=\int \mathcal{D} \mathcal{G} \mathcal{D} \Sigma e^{-N L S_{0 }[\mathcal{G}, \Sigma]-N L S_{I }[\mathcal{G}, \Sigma]}
\label{disorder averaged partition function}
\end{equation}
where the factor of $L$ comes from summing over $L$-identical dots in the chain. We define the Green's function as
\begin{equation}
\mathcal{G}\left(t_1, t_2\right) \equiv \frac{-1}{N} \sum_{\alpha=1}^N\left\langle\mathcal{T}_{\mathcal{C}} c_{ \alpha}\left(t_1\right) c_{ \alpha}^{\dagger}\left(t_2\right)\right\rangle .
\label{green function definition}
\end{equation}

The effective action $S_{\text{eff.}} = S_0 + S_I$ is given by (we provide expressions for the equivalent dot model as in Eq. \eqref{model equivalent equation} and the corresponding expressions for the chain as in Eq. \eqref{model equation} can be found in Eqs. 7, 8, 9 and 10 in Ref. \cite{Jha2023})
\begin{widetext}
    \begin{equation}
S_{0 }=-\operatorname{Tr} \ln \left(\mathcal{G}_{0}^{-1}-\Sigma \right)-\int d t_1 d t_2\left(\Sigma \left(t_1, t_2\right) \mathcal{G} \left(t_2, t_1\right)-\frac{J_q^2}{2 q^2}\left(-4 \mathcal{G}\left(t_1, t_2\right) \mathcal{G} \left(t_2, t_1\right)\right)^{\frac{q}{2}}\right)
\end{equation}
\end{widetext}
where $\Gg_0$ is the free fermionic Green's function as well as
\begin{equation}
S_{I } \equiv \int d t_1 d t_2 \mathscr{L}_{I }[\mathcal{G}]\left(t_1, t_2\right)
\label{interacting action}
\end{equation}
where (corresponding to the dot in Eq. \eqref{model equivalent equation})
\begin{equation}
\mathscr{L}_{I }[\mathcal{G}]=  \frac{K_{\kappa q}^2}{ \kappa q^2 }\left[-4 \mathcal{G} \left(t_1, t_2\right) \mathcal{G}\left(t_2, t_1\right)\right]^{\kappa q / 2} 
\end{equation}
In terms of the chain as in Eq. \eqref{model equation}, the full expression is given in Eq. 10 of Ref. \cite{Jha2023}.

We are employing the real time (Keldysh) formalism, so we can divide the Green's function defined in Eq. \eqref{green function definition} into greater $(t_1>t_2)$ and lesser $(t_1<t_2)$ Green's functions that denote the time ordering of operators on the Keldysh contour (see Fig. \ref{keldysh plane}) as follows:
\be
\Gg(t_1,t_2) = \Theta_{\Cc}(t_1-t_2)\Gg^>(t_1,t_2)+\Theta_{\Cc}(t_2-t_1)\Gg^<(t_1,t_2)
\label{green's function in terms of forward and backward}
\ee
where
\begin{equation}
    \Theta_{\Cc}(t_1-t_2) = \begin{cases} 1 & \text{if } t_1 >_\Cc t_2\\
    0 & \text{if } t_1 <_\Cc t_2
    \end{cases}
    \label{theta function definition}
\end{equation}
In other words, if $t_1,t_2 \in \Cc_-$, then this is just the ordinary theta function. If $t_1,t_2 \in \Cc_+$, then we have $\Theta_{\Cc}(t_1-t_2) = \Theta(t_2-t_1)$. Here, $\Cc_-$ and $\Cc_+$ are the backward and the forward Keldysh contours as shown in Fig. \ref{keldysh plane}. Moreover, we use the large-$q$ ansatz for the total Green's function \cite{Maldacena2016Nov, Louw2022, Louw2023Feb, Jha2023} that leads us to
\begin{equation}
    \Gg^{\gtrless}(t_1, t_2) = \Gg_0^{\gtrless} e^{g^{\gtrless}(t_1, t_2)/q}= \left( \Qq\mp \frac{1}{2}\right) e^{g^{\gtrless}(t_1, t_2)/q}
    \label{ansatz equation}
\end{equation}
with the boundary condition $g^{\gtrless}(t,t) = 0$ and the charge scaling $\Qq = \Oo(q^{-1/2})$. We further introduce the ``symmetric'' and ``antisymmetric'' Green's function as
    \begin{equation}
g^{ \pm}\left(t_1, t_2\right) \equiv \frac{g^{>}\left(t_1, t_2\right) \pm g^{<}\left(t_2, t_1\right)}{2}
\label{g plus minus def}
\end{equation}
where we notice that the time ordering is swapped in the last term on the right-hand side. Note that $g^{\gtrless}\left(t_1, t_2\right)^*=g^{\gtrless}\left(t_2, t_1\right)$. For Majorana case, $g^-(t_1, t_2) = 0$ as it measures fluctuations away from the half-filling case. Therefore, solving for $g^+$ and $g^-$ is equivalent to solving the total Green's function $\Gg^{\gtrless}$ of the system.

Having obtained the effective action, we use the Euler-Lagrange equations to get equations of motion for $\Gg$ and $\Sigma$ which are the Schwinger-Dyson equations. We are interested in equilibrium situation; therefore the functions are time-translational invariant. The Dyson equation connects both and is given by
\begin{equation}
    \Gg_0^{-1} - \Gg^{-1} = \Sigma
    \label{dyson equation}
\end{equation}
where the total self-energy is simply the sum of individual terms $\Sigma = \Sigma_{q} + \Sigma_{\kappa q}$ \cite{Maldacena2016Nov}. To leading order in $1/q$, the self-energies are given by
\begin{equation}
\begin{aligned}
\Sigma_{q}^{\gtrless}\left(t_1, t_2\right)&=\frac{1}{q} \tilde{\mathcal{L}}^{\gtrless}_{ q}\left(t_1, t_2\right) \mathcal{G}^{\gtrless}\left(t_1, t_2\right) + \Oo(q^{-2}) \\
    \Sigma_{\kappa q}^{\gtrless}\left(t_1, t_2\right)&=\frac{1}{q} \mathcal{L}^{\gtrless}_{\kappa q}\left(t_1, t_2\right) \mathcal{G}^{\gtrless}\left(t_1, t_2\right)+ \Oo(q^{-2})
\label{self-energies}
\end{aligned}
\end{equation}
where 
\begin{equation}
\begin{aligned}
\tilde{\mathcal{L}}^{>}_{ q}\left(t_1, t_2\right) &\equiv  2 \mathcal{J}_q^2 e^{ g_{+}\left(t_1, t_2\right)}, \tilde{\mathcal{L}}^{<}\left(t_1, t_2\right)=\tilde{\mathcal{L}}^{>}\left(t_1, t_2\right)^* \\
\mathcal{L}^{>}_{\kappa q}\left(t_1, t_2\right) &\equiv  2 \mathcal{K}_{\kappa q}^2 e^{\kappa g_{+}\left(t_1, t_2\right)}, \mathcal{L}^{<}\left(t_1, t_2\right)=\mathcal{L}^{>}\left(t_1, t_2\right)^*
\end{aligned}
\end{equation}
and 
\begin{equation}
\begin{aligned}
\mathcal{J}_{ q} &\equiv\left(1-4 \mathcal{Q}^2\right)^{ q / 4-1 / 2} K_{ q},\\
\mathcal{K}_{\kappa q} &\equiv\left(1-4 \mathcal{Q}^2\right)^{\kappa q / 4-1 / 2} K_{\kappa q} .
\end{aligned}
\label{curly J definition}
\end{equation}
We will only have non-trivial interaction when $\mathcal{J}_{\kappa q}$ is non-zero in the large-$q$ limit, which we ensure by considering small charge densities which scale like $\Qq = \Oo(q^{-1/2})$.

Therefore, using the ansatz in Eq. \eqref{ansatz equation} in the Schwinger-Dyson equations, we are led to the differential equations for $g^{\pm}$ where the precise form of the differential equation and solution for $g^+$ will depend on the value of $\kappa$ while the differential equation for $g^-$ is given by 
\begin{equation}
         \dot{g}^{-}\left(t\right)=2 \imath \mathcal{Q}  \alpha(t) 
         \label{g-minus differential equation}
\end{equation}
where $\alpha$ depends on $\kappa$ and is related to the expectation values of the Hamiltonian as
\begin{equation}
\left(1-\mathcal{Q}^2\right) \alpha(t)=\epsilon_q(t) + \kappa \epsilon_{\kappa q }(t) 
\label{alpha definition}
\end{equation}
at leading order in $1/q$ and $\epsilon_{\kappa q}$ is given by
\begin{equation}
\epsilon_{ q}(t) \equiv q^2 \frac{J_{ q}\langle\hat{\mathcal{H}}_{ q}\rangle}{N}, \quad \epsilon_{\kappa q}(t) \equiv q^2 \frac{K_{\kappa q}\langle\hat{\mathcal{H}}_{\kappa q}\rangle}{N}.
\label{epsilon definition}
\end{equation}
The $\alpha$ also provides the initial condition for the first derivative of the ``symmetric'' Green's function (see, e.g., Appendix B of Ref. \cite{Louw2023Dec}) 
\begin{equation}
\dot{g}^{+}(0)=\imath \alpha(0)
\label{g-plus initial condition}
\end{equation}
where $\alpha$ is a constant in equilibrium for all time due to energy conservation.

The solution at equilibrium for $g^-(t)$ is given by
\begin{equation}
    g^-(t) = 2 \imath \Qq \alpha  t
    \label{g-minus solution}
\end{equation}
Recall that $\Qq = \Oo(q^{-1/2})$. We are interested in three cases in this work, namely $\kappa = \{\frac{1}{2}, 1, 2 \}$ and we show that for each of these cases, the differential equations in real time for $g^{+}(t)$ are exactly solvable in closed form too just like it's for $g^-(t)$ as shown above, or in other words, the system is analytically solvable in leading order in $1/q$.

\subsection{\texorpdfstring{$\kappa = 1/2$}{}}

The differential equation is given by
\begin{equation}
\ddot{g}^{+}(t)=-2 \Jj_q^2 e^{g^{+}(t)}-2\Kk_{ q/2}^2 e^{g^{+}(t)/2}
\label{k=1/2 differential equation}
\end{equation}
which can be solved exactly for $g^+(t)$ to get
\begin{equation}
e^{g^{+}(t) / 2}=\frac{1}{\left(\beta \mathcal{K}_{q / 2}\right)^2} \frac{(\pi v)^2}{1+\sqrt{A^2+1} \cos (\pi v(1 / 2-\imath t / \beta))}
\label{k=1/2 solution}
\end{equation}
where we define 
\begin{equation}
A \equiv \frac{\pi v \beta \mathcal{J}_q}{\left(\beta \mathcal{K}_{q / 2}\right)^2}.
\label{A definition}
\end{equation}

Using the initial condition for $g^+(t=0) = 0$, we get the closure relation for $v$
\begin{equation}
\pi v=\sqrt{\left(\beta \mathcal{J}_q\right)^2+\left(\frac{\left(\beta \mathcal{K}_{q / 2}\right)^2}{\pi v}\right)^2} \cos (\pi v / 2)+\frac{\left(\beta \mathcal{K}_{q / 2}\right)^2}{\pi v}
\label{k=1/2 closure relation}
\end{equation}

\subsection{\texorpdfstring{$\kappa = 1$}{}}

This reduces to the Hamiltonian in Eq. \eqref{model equivalent equation} as two copies of single large-$q$ complex SYK models. The differential equation becomes
\begin{equation}
\ddot{g}^{+}(t)=-\left(2 \Jj_q^2  + 2 \Kk_q^2 \right)e^{g^{+}(t)},
\label{k=1 differential equation}
\end{equation}
which can be solved exactly for $g^+(t)$ to get
\begin{equation}
e^{g^+(t)} = \frac{(\pi v)^2}{\beta^2(\Jj_q^2 + \Kk_q^2) \cos^2(\pi v(1/2 - \i t/\beta))}
\label{k=1 solution}
\end{equation}

Again, the initial condition for $g^+(t=0)=0$ leads to the following closure relation:
\begin{equation}
\pi v= \beta \sqrt{ \Jj_q^2  +  \Kk_q^2 } \cos (\pi v / 2) .
\label{k=1 closure relation}
\end{equation}

\subsection{\texorpdfstring{$\kappa = 2$}{}}

This case is the reversed situation for $\kappa = 1/2$ where we have to make the following substitutions in $\kappa=1/2$ case: $(g^+, \Kk_{q/2}^2, \Jj_q^2) \to 2 \left(g^+, \Jj_q^2, \Kk_{2q}^2 \right)$ to get the differential equation 
\begin{equation}
 \ddot{g}^{+}(t) = -2 \Kk_{2 q}^2 e^{2 g^{+}(t)} - 2 \Jj_q^2 e^{g^{+}(t) },
 \label{k=2 differential equation}
\end{equation}
which can be solved exactly for $g^+(t)$ to get
\begin{equation}
e^{g^{+}(t) }=\frac{1}{2\left( \beta \mathcal{J}_{q }\right)^2} \frac{(\pi v)^2}{1+\sqrt{B^2+1} \cos (\pi v(1 / 2-\imath t / \beta))}
\label{k=2 solution}
\end{equation}
where we define
\begin{equation}
B \equiv \frac{ \pi v \beta \mathcal{K}_{2q}}{\sqrt{2} \left( \beta \mathcal{J}_{q }\right)^2}.
\label{B definition}
\end{equation}

The closure relation is given by
\begin{equation}
\pi v=\sqrt{2 \left( \beta \mathcal{K}_{2q}\right)^2+\left(\frac{2\left( \beta \mathcal{J}_{q}\right)^2}{\pi v}\right)^2} \cos (\pi v / 2)+\frac{2 \left( \beta \mathcal{J}_{q }\right)^2}{\pi v}
\label{k=2 closure relation}
\end{equation}

We will see later that although $\kappa =2$ case is related by a simple transformation as mentioned above to the $\kappa = 1/2$ case, the physics behind the two cases differ drastically when it comes to the transport properties, namely $\kappa = 2$ has an insulator resistivity while $\kappa = 1/2$ is a conductor.

\subsection{A note on scaling arguments}
\label{a note on scaling arguments}

As temperature tends to zero, we know from naive scaling arguments of interacting fermions \cite{Shankar1994Jan} that the term in the Hamiltonian containing fewer number of fermions tend to dominate the overall behavior of the system. This would \textit{suggest} that in our case; therefore, the hopping term dominates for $\kappa = 1/2$ case while on-site SYK term dominates for $\kappa = 2$ case. This results in an conducting as well as insulating behavior at low temperatures, respectively. For the $\kappa = 1$ case, both the on-site as well as the hopping terms have the same number of fermions and therefore both compete, and we cannot make an \textit{a priori} comment on the transport property from the scaling arguments. However, there is another naive argument from the renormalization group (RG) that since the fermionic fields have zero mass-dimensions in $0+1$-dimension (explained below), then the coupling constants appearing in the Hamiltonian in Eq. \eqref{model equivalent equation} have the same mass dimensions, namely $[J_q] = [K_{\kappa q}]$. So a naive application of RG arguments would then imply that all of these terms have the same relevance in all the three cases ($\kappa = \{1/2,1,2\}$). However, this is also not true as is explained in this section. 

We now present some scaling arguments to make our discussion precise. We start with a single large-$q$ complex SYK model. We know that the action has to be dimensionless as it goes in the exponential, then in $0+1$-dimension, the fermionic field can be shown to have zero mass dimensions. This can be seen, in general, using the dimensional analysis of a $d+1$-dimensional Hamiltonian which implies that the fermions will have a mass dimension of $d/2$, as seen from the kinetic part of the action having to be dimensionless $S_0 = \int d^{d+1} x c^\dag \p_t c$. Furthermore, the conformal limit solution for the Green's function goes as $\Gg_c(t) \sim |t|^{-2/q}$ which implies that the scaling dimension of the fermionic field is $1/q$ which is the same as the anomalous dimension (since mass dimension vanishes in $0+1$-dimension for fermionic fields and anomalous dimension is given by the difference between the scaling dimension and the mass dimension). So each SYK$_{\kappa q}$ individually has a fermionic field with a scaling dimension of $1/(\kappa q)$. However, to get the combined scaling dimension, one can look at the conformal solution $\Gg_c^\kappa (t) \sim t^{-2\Delta_\kappa}$ where the scaling dimension can be read-off as $\Delta_\kappa$ (recall $\Gg(t) \propto \langle c^\dag_i(t) c_i\rangle$). Since we know the exact solutions for all the three models we have considered (Eqs. \eqref{k=1/2 solution}, \eqref{k=1 solution} and \eqref{k=2 solution}), all we have to do is to take the conformal limit that emerges in the zero-temperature limit and get the scaling dimension.

Since we have a zero spatial dimensional dot, the mass dimension of the fermionic field is zero. From this, it would imply that the coupling constants for all the three cases have a dimension of mass, i.e., $[J_q] = [K_{\kappa q}]$ in Eq. \eqref{model equivalent equation} that would suggest that both the terms have the same relevance. As we will see now, this is only true for $\kappa = 1$ case. For $\kappa = 1$, we know that the zero-temperature behavior is obtained by taking the zero-temperature limit of Eq. \eqref{k=1 solution} which leads to
\be
\Gg_c^{\kappa = 1}(t) \sim |t|^{-2/q}.
\ee
Therefore, a true scaling dimension is $\Delta_{\kappa = 1} = 1/q$ which is the same as the anomalous dimension \cite{Polchinski2016Apr}. 

Now we consider the $\kappa =2$ case where we have two competing terms with different scaling dimensions. The question of what the combined scaling dimension becomes is addressed by considering the Green's function in zero-temperature limit for this model. In this case, close to zero temperature, what we find from Eq. \eqref{k=2 solution} is
\begin{equation}
e^{g^{+}(t)}\sim \frac{1}{1+ \cos (\pi v(1 / 2-\imath t / \beta))} 
\end{equation}
which behaves as the square root of the $\kappa = 1$ Green's function (Eq. \eqref{k=1 solution}). Therefore, $\Gg_c^{\kappa =2}(t) \sim |t|^{-1/q}$ which implies a scaling dimension of $\Delta_{\kappa = 2} = 1/(2q)$ which is less than $1/q$. Therefore the $SYK_q$ term dominates. Similarly for $\kappa = 1/2$, we find $\Gg_c^{\kappa =1/2}(t) \sim |t|^{-4/q}$, hence the scaling dimension becomes $\Delta_{\kappa =1/2} = 2/q$ which is greater than $1/q$. This implies that the $\text{SYK}_{q/2}$ term dominates. In general, we can combined the three cases by noting that the scaling dimension is $\Delta_\kappa = 1/\text{min}\{ \kappa q, q\}$.

Having said that, the scaling arguments are not rigorous and there exists no proof away from the critical points for SYK-like systems. Therefore, scaling arguments are not a substitute for a rigorous mathematical treatment of the systems that we are considering. Using the benefit of hindsight where we have obtained rigorous mathematical results about transport properties for the aforementioned three cases, we confirm the aforementioned scaling arguments for $\kappa = 1/2$ as well as $\kappa = 2$ cases that indeed, they have conducting and insulating behaviors at low temperatures, respectively. In addition, we also find a conducting behavior for the $\kappa = 1$ case which cannot be addressed from the perspective of scaling arguments, as both the on-site as well as the hopping terms compete against each other as temperature is reduced to zero. We now embark on a detailed mathematical journey illustrating the transport properties for these three cases and illustrate that the naive expectations from scaling arguments hold for $\kappa = \{1/2,2\}$ cases.

\section{Keldysh Contour Deformations}
\label{keldysh contour deformation section}

We develop a general methodology for SYK-like systems using the Keldysh formalism \cite{Stefanucci2013Mar} and use the formalism to calculate current-current correlations and conductivity using linear response theory. We start with deriving the general relations for contour deformations in the Keldysh plane for SYK-like systems that will allow us to calculate expectation values of various quantities.

\begin{figure}[ht]
	\centering
	\captionsetup{justification=centering}
	\includegraphics[width=\columnwidth]{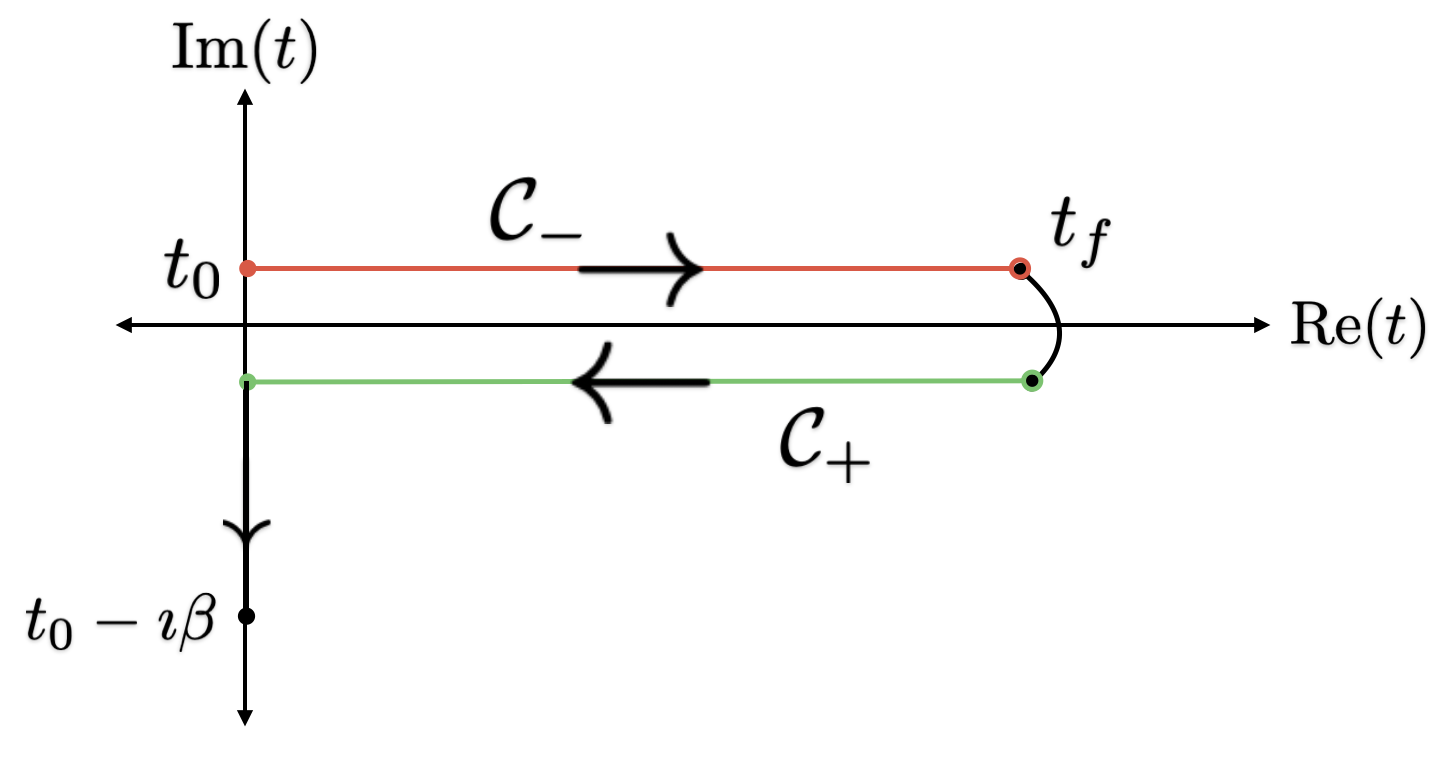}
	\caption{Keldysh contour}
	\label{keldysh plane}
\end{figure}

\subsection{Time evolution in Keldysh plane}
\label{time evolution subsection}

We wish to show how one can derive various time-dependent expectation values from the Keldysh action, or rather the Keldysh generating function $\Zz$. This function $\Zz$ is the trace of the unitary operator given by
\be 
\Zz = \Tr[U_\Cc] \qquad \text{where} \quad U_\Cc = \Tt \exp\left(-\i \nint[\Cc]{t} \lambda(t)\hat{\Hh}(t)\right)
\ee
where $\Cc$ is the full Keldysh contour going from $t_0$ to $\infty$ and then back to $t_0$ as shown in Fig. \ref{keldysh plane}. The upper (forward) contour is denoted by $\Cc_-$ and the lower (backward) contour is denoted by $\Cc_+$. Note that this operator has both the implicit time dependence stemming from the coupling $\lambda(t)$ and the explicit time dependence coming from
\begin{equation}
	\hat{\Hh}(\tau) = U(t_0,\tau)\hat{\Hh}(t_0)U(t_0,\tau)^\dag.
\end{equation}
Only the implicit time dependence  $\lambda(t)\hat{\Hh}(t_0)$ is important for the unitary time evolution operator.

We can separate $U_\Cc$ over the forward $\Cc_-$ and the backward $\Cc_+$ contours, where (for small time steps $\delta t$)
\begin{subequations}
	\begin{align}
		U_{\Cc_-} &= (1-\i \delta t \hat{\Hh}(t_0))(1-\i \delta t \hat{\Hh}(t_0+\delta t))\cdots (1-\i \delta t \hat{\Hh}(t_f)) \\
		U_{\Cc_+} &= (1+\i \delta t \hat{\Hh}(t_f))(1+\i \delta t \hat{\Hh}(t_f+\delta t))\cdots (1+\i \delta t \hat{\Hh}(t_0)) 
	\end{align}
\end{subequations}

Since we are in total evolving both forward and backward in time, we actually end up with $U_\Cc = 1$. To still get out expectation values, we illustrate here the forward contour deformation by changing the couplings to be contour dependent as
\be D_{-}(t) = D(t)+  \Delta(t),\quad D_{+}(t) = D(t) \label{coupling def along contour} \ee
where we put potential on the forward contour $\Cc_-$
\be \Delta(t) = \sum_{n}\ll_n D(\tau_n)\delta(\tau_n-t). \label{Delta def} \ee 
Note that this is an example for the forward contour deformation. We will generalize this to the most general forward and backward contour deformations below and derive general relations to be used later in this manuscript. 

Since the above operator has a time-dependent coupling, we can extract out the following for a particular value of $\tau$ corresponding to a particular $\ll$:
\begin{widetext}
\begin{equation}
\begin{aligned}
	\frac{\delta}{\delta \ll(\tau)} U_{\Cc_-} &= \lim_{\l \to 0}\frac{U_{\Cc_-}[D(t)+\ll D(\tau)\delta(\tau-t)]-U_{\Cc_-}[D(t)]}{\ll} = U(t_0,\tau) (-\i \hat{\Hh}(t_0))U(\tau,t_f) \\
	&= U(t_0,\tau) (-\i U(t_0,\tau)^\dag \hat{\Hh}(\tau) U(t_0,\tau) )U(\tau,t_f) 
	= (-\i \hat{\Hh}(\tau))U(t_0,t_f) \label{expecEg}
\end{aligned}
\end{equation}
\end{widetext}
Note that $U_{\Cc_+}$ does not have a $\ll$ dependence. The above functional derivative (Eq. \eqref{expecEg}) then corresponds to a partial derivative with respect to $\ll$ where $U_{\Cc_+}$ adds another time evolution from $t_f$ to $t_0$ canceling the last term in Eq. \eqref{expecEg} to get for the full-time evolution operator
\be 
\begin{aligned}
\p_{\ll_n} U_{\Cc}\vert_{\vec{\ll} = 0} &= U(t_0,\tau_n) (-\i \hat{\Hh}(t_0))U(\tau_n,t_f)U(t_f,t_0) \\
&=  -\i \hat{\Hh}(\tau_n) 
\end{aligned}
\ee
As such, the energy at a time $\tau_n$ is given by
\begin{equation}
\ex{\hat{\Hh}(\tau_n)}(\tau_n) = \i \p_{\ll_n} \Zz  \vert_{\vec{\ll} = 0}
\label{expectation of H}
\end{equation}
The resulting expectation value on the left-hand side can independently be reproduced using the Galitskii-Migdal relation \cite{Louw2022}, provided that the Green's functions are known where we should find a negative energy and a positive conductivity.


\subsection{Partition function}
	\label{partition function subsection}

The (disorder-averaged) generating function is given in Eq. \eqref{disorder averaged partition function}. The expression for the interacting part $S_I$ (as given in Eq. \eqref{interacting action}) is re-written in the following general form (recall that the coupling constants $\{ J, D \}$ in the chain in Eq. \eqref{model equation} is mapped to the coupling constants $\{ J_q, J_{\kappa q} \}$ of the dot in Eq. \eqref{model equivalent equation}):
\begin{equation}
	S_{I}[\Gg] \equiv \nint[\Cc]{t_1} \nint[\Cc]{t_2}\frac{D_{\Cc}(t_1)D_{\Cc}(t_2)}{2}  F[\Gg(t_1,t_2)\Gg(t_2,t_1)] 
	\label{interacting action general form}
\end{equation}
Note that we are considering the general non-equilibrium (non-time-translation invariant quantities) case where we are going to derive generalized results that we will use to address our equilibrium situation later in this work. Here $S_{I}$ is the only coupling dependent part (we assume the coupling constant is real) where $F$ is some general functional depending only on a single argument, namely $\Gg(t_1,t_2)\Gg(t_2,t_1)$. Moreover, $\Cc$ denotes the Keldysh contour in the Keldysh plane, as shown in Fig. \ref{keldysh plane}. The coupling constant along the contours is denoted as (using the same approach as in Eq. \eqref{coupling def along contour})
\begin{equation}
	D(t)|_{\Cc_-} \equiv D_-(t) \qquad 	D(t)|_{\Cc_+} \equiv D_+(t) 
	\label{J+ and J- def}
\end{equation} 
such that
\begin{equation}
	D(t)|_{\Cc_-} - 	D(t)|_{\Cc_+}=    D_-(t) -  D_+(t) \equiv \Delta (t)
	\label{delta def}
\end{equation}

To simplify matters, we wish to express our functional $F$ in terms of $\Gg^>$ and $\Gg^<$ using Eq. \eqref{green's function in terms of forward and backward} and expand as
\begin{equation}
\begin{aligned}
	\Gg(t_1,t_2)\Gg(t_2,t_1) = &\Theta_{\Cc}(t_1-t_2)\Gg^>(t_1,t_2)\Gg^<(t_2,t_1)\\
 &+\Theta_{\Cc}(t_2-t_1)\left[\Gg^>(t_1,t_2)\Gg^<(t_2,t_1)\right]^*
 \end{aligned}
\end{equation}
where we used the fact that
\be
\left[\Gg^>(t_1,t_2)\Gg^<(t_2,t_1)\right]^* = \Gg^>(t_2,t_1)\Gg^<(t_1,t_2) 
\label{conjugate green function}
\ee
Then we have our functional $F$ written as
\be
\begin{aligned}
F[\Gg(t_1,t_2)\Gg(t_2,t_1)]= &\Theta_{\Cc}(t_1-t_2)F[\Gg^>(t_1,t_2)\Gg^<(t_2,t_1)]\\
&+\Theta_{\Cc}(t_2-t_1)F^*[\Gg^>(t_1,t_2)\Gg^<(t_2,t_1)]
\label{F in terms of G and G lesser and greater}
\end{aligned}
\ee

To simplify further, we separate $F[\Gg^>(t_1,t_2)\Gg^<(t_2,t_1)]$ into real and imaginary parts as
\begin{equation}
	F[\Gg^>(t_1,t_2)\Gg^<(t_2,t_1)] \equiv X(t_1,t_2) + \i Y(t_1,t_2) = F(t_1, t_2)
	\label{X and Y def}
\end{equation}
where $X(t_1,t_2)$ and $Y(t_1,t_2)$ are real functions, and we have chosen in the last equality to show the implicit dependency of our functional $F$ on time arguments $t_1$ and $t_2$. Taking complex conjugate on both sided and using Eq. \eqref{conjugate green function}, we have
\begin{subequations}
    	\begin{align}
		F^*[\Gg^>(t_1,t_2)\Gg^<(t_2,t_1)] &= F[\Gg^>(t_2,t_1)\Gg^<(t_1,t_2)] \\
		\left[X(t_1,t_2) + \i Y(t_1,t_2) \right]^* &= X(t_1,t_2) - \i Y(t_1,t_2)
	\end{align}
 \end{subequations}
Therefore comparing against Eq. \eqref{X and Y def}, we get
\begin{equation}
	X(t_1,t_2) = X(t_2,t_1) \qquad Y(t_1,t_2) = - Y(t_2,t_1)
	\label{conditions on X and Y}
\end{equation}
Plugging Eq. \eqref{conditions on X and Y} into Eq. \eqref{F in terms of G and G lesser and greater}, we get
\begin{widetext}
\be
F[\Gg(t_1,t_2)\Gg(t_2,t_1)]= \left[ \Theta_{\Cc}(t_1-t_2) + \Theta_{\Cc}(t_2-t_1) \right] X(t_1, t_2) + \i \left[ \Theta_{\Cc}(t_1-t_2) - \Theta_{\Cc}(t_2-t_1) \right] Y(t_1, t_2) 
\ee
\end{widetext}
Therefore we are finally left with
\be
F[\Gg(t_1,t_2)\Gg(t_2,t_1)] =  X(t_1, t_2)  + \sgn_{\Cc}(t_1-t_2) \i Y(t_1,t_2)
\label{F in terms of X and Y}
\ee
We have the mathematical identity $\sgn_\Cc (t_1 - t_2) = 2 \Theta_\Cc(t_1 - t_2) - 1$ which translates to the following for various parts of the Keldysh contour:
\begin{equation}
    \sgn_\Cc(t_1 - t_2) = \begin{cases}  \sgn(t_1 - t_2)& \text{if } t_1,t_2 \in \Cc_-\\
\sgn(t_2 - t_1)& \text{if } t_1,t_2 \in \Cc_+\\
-1 & \text{if } t_1 \in \Cc_-,t_2 \in \Cc_+ \\
+1 & \text{if } t_1 \in \Cc_+,t_2 \in \Cc_-
    \end{cases}
    \label{signum}
\end{equation}

Thus to summarize, the coupling dependency in action solely appears in the interacting action given in Eq. \eqref{interacting action general form} which depends on some generic $F$ that is a functional of $\Gg(t_1,t_2)\Gg(t_2,t_1)$. We re-expressed this $F[\Gg(t_1,t_2)\Gg(t_2,t_1)]$ in terms of $X(t_1, t_2)$ and $Y(t_1, t_2)$ as in Eq. \eqref{F in terms of X and Y} where $X(t_1, t_2)$ and $Y(t_1, t_2)$ (both being real functions) are real and imaginary parts of $F[\Gg^>(t_1,t_2)\Gg^<(t_2,t_1)] $ as in Eq. \eqref{X and Y def} satisfying the property in Eq. \eqref{conditions on X and Y}.  

\subsection{Forward \& Backward Contour Deformations and Interacting Action}

We now go to the most general case, where we deform both the forward and the backward Keldysh contour in Fig. \ref{keldysh plane}. We also assume general coupling constants that can take complex values and consider the non-equilibrium situation. Having derived the most general relations, we then employ those results for our equilibrium situation with real coupling constants. 

We start with generalizing the form of the interacting action (complex-valued coupling constants) as given in Eq. \eqref{interacting action general form} 
\begin{widetext}
\begin{equation}
	\begin{aligned}
		S_{I}[\Gg] &=  \nint[\Cc]{t_1} \nint[\Cc]{t_2}\frac{D_{\Cc}(t_1)\bar{D}_{\Cc}(t_2) }{2}  F(\Gg(t_1,t_2)\Gg(t_2,t_1))   = \nint[\Cc]{t_1} \nint[\Cc]{t_2}\frac{D_{\Cc}(t_1)\bar{D}_{\Cc}(t_2) }{2}  \left( X (t_1, t_2) +\sgn_\Cc(t_1 - t_2) \i Y(t_1, t_2) \right) \\
		&= \Re S_I + \i \Im S_I \qquad \left(X(t_1, t_2) = X(t_2, t_1) \text{ and } Y(t_1, t_2) = - Y(t_2, t_1)\right)
	\end{aligned}
 \label{interacting action general complex-valued form}
\end{equation}
\end{widetext}
where $D_\Cc(t) = D_-(t) + D_+(t)$ and $\bar{D}_\Cc(t) = \bar{D}_-(t) + \bar{D}_+(t)$ are complex-valued coupling constant and its complex conjugate, respectively.

We now deform both the forward and the backward contours as follows:
\begin{subequations}
\begin{align}
    D_-(t) &= D(t) + \Delta_-(t) = D(t)+ \sum_n \ll_n D(\tau_n) \delta(\tau_n - t), \\
    \bar{D}_-(t) &= D^*(t) + \bar{\Delta}_-(t) =  D^*(t) + \sum_n \llbar_n D^*(\tau_n) \delta(\tau_n - t) , \\
    D_+(t) &= D(t)- \Delta_+(t) = D(t) - \sum_n \et_n D(\tau_n) \delta(\tau_n - t), \\
    \bar{D}_+(t) &= D^*(t) - \bar{\Delta}_+(t)= D^*(t) - \sum_n \etbar_n D^*(\tau_n) \delta(\tau_n - t) 
    \end{align}
    \label{deformation contours back and forth}
\end{subequations}

We begin by first considering the real part of the interacting action. Before we proceed, we remind that the Keldysh contour along $\Cc_-$ runs from $t_0$ to $t_f$ while the contour $\Cc_+$ runs in reverse from $t_f$ to $t_0$ which we consider with a minus sign as follows:
\begin{subequations}
	\begin{align}
		\nint[\Cc_-][]{t} (\cdots) &= \nint[t_0][t_f]{t} (\cdots)  \\
		\nint[\Cc_+][]{t} (\cdots) &= \nint[t_f][t_0]{t} (\cdots) = -\nint[t_0][t_f]{t} (\cdots) 
	\end{align}
\end{subequations}
The general results for the real and imaginary parts of the interacting action is calculated in detail in Appendix \ref{app. forward and backward contour deformation} and is provided in Eqs. \eqref{real action total 2} and \eqref{imaginary action total 2}, respectively. Since the total interacting action is a sum of real and imaginary parts, the full expression is provided in Eq. \eqref{full interacting action in terms of lambda and eta}. Then we employ these results for our equilibrium case where we assume time-independent couplings and initial time $t_0 \to -\infty$ to get 
\begin{widetext}
    \begin{equation}
    \begin{aligned}
        \frac{S_I}{|D|^2}  =& S_I |_{\vec{\ll}, \vec{\et} = 0} + \frac{1}{2} \sum\limits_{n>m} \left[ (\ll_n + \et_n) (\llbar_m F(\tau_n - \tau_m) + \etbar_m F^*(\tau_n - \tau_m) ) + (\llbar_n + \etbar_n)(\ll_m F(\tau_n - \tau_m)  +  \et_m F^*(\tau_n - \tau_m)) \right]  \\
        &- \i \sum_m \left[  \llbar_m + \ll_m + \et_m +  \etbar_m  \right] \left( \nint[-\infty][0]{t}  Y(t) \right) +\frac{1}{2} \sum\limits_{n} (\ll_n + \et_n)(\llbar_n + \etbar_n) X(0)  \qquad (\text{Equilibrium})
    \end{aligned}
    \label{equilibrium effective action in terms of lambda and eta}
\end{equation}
\end{widetext}
where $\vec{\ll} = \left( \{\ll_n \}, \{\llbar_m \} \right)$ and $\vec{\et} = \left( \{\et_n \}, \{\etbar_m \} \right)$.

We can now specialize to our case, where $S_I$ is given in Eq. \eqref{interacting action}. We use the large-$q$ ansatz for the Green's function in Eq. \eqref{ansatz equation}. Therefore, using Section \ref{partition function subsection}, we have for $F(t)$ the following expression (recall we are in the large-$q$ limit and $\Qq = \Oo(1/\sqrt{q})$): 
\begin{equation}
\begin{aligned}
    F(t) =& \frac{1}{\kappa q^2} \left[ -4 \Gg^>(t)\Gg^<(-t) \right]^{\kappa q/2} \\
    =& \frac{1}{\kappa q^2 } \left[ \left( 1 - 4 \Qq^2 \right) e^{2 g^+(t)/q} \right]^{\kappa q/2}  \\
    =& \frac{1}{\kappa q^2 } \left[e^{-4 \Qq^2} e^{2 g^+(t)/q} \right]^{\kappa q/2} \\
	=& \frac{1}{\kappa q^2 } e^{- 2\kappa q \Qq^2} e^{\kappa g^+(t)} = X(t) + \i Y(t)
\end{aligned}
\label{expression for F}
\end{equation}

Therefore, just like we did in Eq. \eqref{expectation of H} where we deformed only the forward contour, we are in a position to calculate various expectation values that we will heavily use later in this work to calculate the current-current correlations that will intern be used to calculate conductivity using the linear response theory.

\section{Current}
\label{current section}

In this section, we derive relations related to the current that is defined through the continuity equation, as explained below.  We start with the continuity equation, followed by a stability analysis in the form of linear response theory and Kubo's formula to study the current response. This leads us to expressions for dynamical and DC conductivities, which we show are related to the current-current correlations. 

\subsection{Continuity equation and current density}
\label{continuity equation subsection}

We focus on the chain electronic current density, which is calculated by the continuity equation $\p_t \hat{\Qq}_{i} =  -\nabla \hat{j}^\Qq$. In an effective one-dimensional system, this is merely 
\begin{equation}
    \p_t \hat{\Qq}_{i}(t) = -\p_x \hat{j}^{\Qq}(x_i,t) = -\frac{\hat{j}^\Qq(x_i+dx,t)-\hat{j}^\Qq(x_i,t)}{dx}.  \label{contEq}
\end{equation}
Here $\hat{j}^\Qq(x_i)$ is the local current density (per flavor) operator, $x_i = i/L$ and $dx = 1/L$,  and $\hat{\Qq}_i$ is the local charge per flavor (also sometimes called the local charge density). Both are intensive in both $N$. From the Heisenberg equation of motion, we find the right-hand side must be equal to $\i[\hat{\Qq}_{i}(t),\hat{\Hh}]$. 

For this section, we define
\begin{equation}
    \hat{\Hh}^{i,i+1} \equiv \hat{\Hh}_{i \to i+1} + \hat{\Hh}_{i \to i+1}^\dagger
    \label{H i, i+1 defined}
\end{equation}
where we are considering the chain as in Eq. \eqref{model equation}.

For the discrete case, we take $x_i = i/L$, $dx = 1/L$. With this, we have the equation 
\begin{equation}
  \frac{\hat{j}^\Qq(x_i)- \hat{j}^\Qq(x_i+dx)}{dx} = \i [\hat{\Qq}_{i},\hat{\Hh}^{i,i+1}] + \i [\hat{\Qq}_{i},\hat{\Hh}^{i-1,i}]
\end{equation}
A natural claim seems to be that $\hat{j}^\Qq_{x}/dx = \i [\hat{\Qq}_{i},\hat{\Hh}^{i-1,i}]$. To show this, we must prove that 
\begin{equation}
    \hat{j}^\Qq(x+dx)/dx = -\i [\hat{\Qq}_{i},\hat{\Hh}^{i,i+1}]  = \i [\hat{\Qq}_{y},\hat{\Hh}^{y-1,y}] (y \equiv i+1). \label{JsymRel}
\end{equation}
For $r$-particle hopping (where $r=\kappa q$ in our case), one may use the generalized Galitskii-Migdal relation to evaluate the above commutator to find \cite{Louw2022}
\begin{equation}
    [\hat{\Qq}_{i},\hat{\Hh}_{i\to i+1}] = \frac{1}{N} \sum_{\alpha}[c_{i;\alpha}^\dag,\hat{\Hh}_{i\to i+1}] c_{i;\alpha} = -\frac{r \hat{\Hh}_{i\to i+1}}{2N} . \label{Galit1}
\end{equation}
With this, we take the conjugate of the relation to get
\begin{equation}
    [\hat{\Qq}_{i},\hat{\Hh}_{i\to i+1}^\dag] = \frac{r}{2N} \hat{\Hh}_{i\to i+1}^\dag. \label{Galit2}
\end{equation}
Therefore we identify with what we are left with as the local extensive (in flavor $N$) current operator
\be \hat{I}_{i+1} = \i r\frac{\hat{\Hh}_{i\to i+1}-\hat{\Hh}_{i\to i+1}^\dag}{2} \label{local current def}\ee
Via a discretization of the continuity equation on a lattice, we identify the total extensive (in $N$ and $L$) current operators as 
     \begin{equation}
        \hat{I} =  \frac{\i r \hat{\Hh}_{\rightarrow}  - \i r  \hat{\Hh}_{\rightarrow}^\dag}{2} \label{currentOp}
    \end{equation}
where $\hat{\Hh}_{\rightarrow}$ is defined in Eq. \eqref{H rightarrow defined}. Therefore, we use the continuity equation to identify the extensive local and total current operators, as in Eqs. \eqref{local current def} and \eqref{currentOp}. As such, we need to extract out $\hat{\Hh}_{\rightarrow}$. We can do this by taking functional derivatives of the action \footnote{Note that in Ref. \cite{Zanoci2022Jun}, they consider the current operator for a particular flavor. This allows the order to be of $\Oo(N^0)$. However, we find this convention of thinking of current as being carried out by different flavors unphysical since there is only a single current corresponding to all flavors. In Refs. \cite{Song2017Nov, Cha2020Sep}, they actually consider the current to be of order $\Oo(N)$ like we do here.}.

\subsection{Current response}
\label{current response subsection}

We consider the case where the system is initially in thermal equilibrium. Let us consider a uniform system weakly perturbed out of equilibrium \footnote{The fact that it is weakly perturbed out of equilibrium will mean that the response function (or conductivity kernel) is only dependent on time differences. The fact that the system has translation invariance, i.e., a periodic crystal, will mean that the conductivity is only dependent on spatial differences.}. A current in the $\alpha$-direction at $(t,r)$ 
     \begin{equation}
         I_{\alpha;\mathbf{x}}(t) = \nint{t'}\frac{1}{L^d}\sum_{\mathbf{x}'\in \{1,2,\ldots,L\}^d} \sum_{\beta = 1}^d\sigma^{\alpha\beta}_{\mathbf{x}-\mathbf{x}'}(t-t') E_{\mathbf{x}'}^\beta(t') \label{currentDef}
     \end{equation}
is the response to some applied electric field $\vec{E}$ in $d$-spatial dimensions. The conductivity tensor $\sigma^{\alpha\beta}$ describes the current response in the direction $\alpha$ to a field in the direction $\beta$.  As such, without any field, there is no charge flow. Due to the system being weakly perturbed out of equilibrium, the conductivity tensor is only dependent on time differences \cite{Bruus2004Sep}. Further, we consider the electric field to be only in the $x_1$-direction. Further, we will suppress the index $\alpha = 1$, so $I_{1;\mathbf{r}}\to I_{x}$ and $\vec{E} = (E,0,0)$. This double convolution may be Fourier transformed to yield $I_{k}(\omega) = \sigma^{11}_k(\omega) E_k(\omega)$.
 
Next we consider a uniform electrical field $\vec{E}(t)$, with magnitude $E(t)$. This also means that $\sigma^{\alpha 1}$ can only depend on distances. For such a field we are left with the Kubo's formula
    \begin{equation} 
        I_{x}(t) = \nint[-\infty][t]{\tau} \sigma(t-\tau) E(\tau), \label{Kubo0}
    \end{equation}
    Taking the Fourier transform, we are left with $I_x(\omega) = E(\omega) \sigma(\omega)$.
   Consider the linear response to a perturbation $\Hh_s[E(t)] = E(t) \hat{X}$. Here the (extensive in $L$ and $N$) polarization operator
\begin{equation}
    \hat{X} = \sum_{j=1}^L j N\hat{\Qq}_{i}
    \label{polarization operator definition}
\end{equation}
as defined in the equation following Eq. (17) of \cite{Takayoshi2022Nov}. It tells us about the distribution of some total charge, hence serves as a sort of position operator. For instance, if we have a total charge $\hat{\Qq}$ and we get $\hat{X}= x_0 \hat{\Qq}$, then we know that all charge is located at the position $x=x_0$. This is due to the ``velocity'' expression 
\begin{equation}
    	\p_t \hat{X}= \i [\Hh, \hat{X}]= N\sum_{j=1}^{L} j \p_t \hat{\Qq}_{j}= \sum_{j=1}^L (j\hat{I}_{j}-j\hat{I}_{j+1}) = \hat{I}
\end{equation}
Here we have defined the total extensive (in $L$ and $N$) current operator $\hat{I} \equiv \sum_{j=1}^L \hat{I}_{j}$.

Here we wish to study the current along the $x$-direction $	\hat{I} = \sum_{x} \hat{I}_{x}$. In general, the response from some observable $\hat{I}$ to such an applied electric field $E(t)$, in the $x$-direction, is given by the Kubo's formula (Eq. \eqref{Kubo0}), with the explicit response function  
\begin{equation}
 \sigma(t) \equiv \Theta(t) \i\ex{[\hat{X}(-t),\hat{I}_x(0)]} = \frac{\Theta(t)}{L}\i\ex{[\hat{X}(-t),\hat{I}(0)]}
 \label{sigma in terms of X commutator}
\end{equation}
where $\hat{X}(t)$ evolves under $\hat{\Hh}$, i.e., the interaction picture. Due to thermal equilibrium (time-translational invariance), we were able to rewrite $\ex{\hat{I}_x(t)\hat{X}(t -\tau)}$ as $\ex{\hat{I}_x(0)\hat{X}(-\tau)}$. We have used the translation invariance to replace $\hat{I}_x(0)$ with $\hat{I}(0)/L$. Again via this invariance, $\sigma_{i} = \sigma$, hence the sum weighted sum over lattice sites is merely $\sigma(\tau)$. With this, the current may be rewritten as 
    \begin{equation}
        I_i(t) = E \nint[0][t]{t'} \sigma(t')
    \end{equation}
with initial conditions for $\sigma(0)$ is derived in Appendix \ref{app. initial condition for sigma} and is given by 
    \begin{equation}
        \sigma(0) =  -\frac{r^2}{4} \frac{\ex{ \Hh_{\text{trans}}}}{L} = 2\, \Im \nint[-\infty][0]{\tau}   \ex{\hat{I}(0) \hat{I}(\tau) }/L
        \label{initial condition for sigma equation}
    \end{equation}
where $\hat{\Hh}_{\text{trans}}$ is defined in Eq. \eqref{H dot and H trans defined}.

Kubo's formula then amounts to a convolution which has the Fourier transform $I_x(\omega) = E(\omega) \sigma(\omega)$, where the term on the right-hand side $\sigma(\omega)$ is the dynamical conductivity. Next, we evaluate $\sigma(\omega)$ using integration by parts, noting its relation to the current-current correlation function as
    \begin{equation}
    \begin{aligned}
	\dot{\sigma}(t) =& \Theta(t)  \frac{\ex{[\hat{I}(-t),\hat{I}(0)]}}{\i L} \\
 =&\Theta(t)  \frac{\ex{[\hat{I}(0),\hat{I}(t)]}}{\i L} = \Theta(t) \frac{2\, \Im\ex{\hat{I}(0)\hat{I}(t)}}{L} 
 \end{aligned}
 \label{sigma dot equation}
    \end{equation}
where we consider without loss of generality that time $t>0$. Here we have used the fact that Hermitian operators $\ex{\hat{I}(0)\hat{I}(t)} =\text{Tr}\{\hat{I}(0)\hat{I}(t)\rho\} = \text{Tr}\{ (\rho \hat{I}(t) \hat{I}(0))^\dag\} = \ex{\hat{I}(t) \hat{I}(0)}^*$ (since density matrices always satisfy $\rho^\dagger = \rho$). We can now integrate $\dot{\sigma}(t)$ where we start the integration from $-\infty$ to account for the initial condition in Eq. \eqref{initial condition for sigma equation} to get
    \begin{equation}
        \sigma(t) = 2\, \Im \nint[-\infty][t]{\tau} \ex{\hat{I}(0)\hat{I}(\tau)}/L.
        \label{dynamical conductivity formula}
    \end{equation}
    We will evaluate the current-current correlation $\ex{\hat{I}(0)\hat{I}(t)}$ later in Section \ref{current correlations section}.

    We are mainly be interested in the DC conductivity. To find this, we take the Fourier transform using integration by parts
\be
\begin{aligned}
\Ff[\sigma](\omega) =& \nint{\tau} e^{\i \omega \tau} \sigma(\tau)= \nint{\tau} \left(\p_{\tau}\frac{e^{\i \omega \tau}}{\i \omega}\right) \sigma(\tau) \\
=& \nint{\tau} \frac{e^{\i \omega \tau}}{\i \omega} \dot{\sigma}(\tau) = \frac{1}{\i \omega} \Ff[\dot{\sigma}](\omega)
\end{aligned}
\label{equation 85}
\ee
as in \cite{Takayoshi2022Nov}. It is typical to denote the above susceptibility by $\chi^R(\tau) \equiv \dot{\sigma}(\tau)$. With this we are left with the formula for the dynamical conductivity \cite{Takayoshi2022Nov} $\Re \Ff[\sigma](\omega) = - \Im\Ff[\dot{\sigma}](\omega)/\omega$.
As derived in Appendix \ref{app. fluctuation dissipation theorem}, the fluctuation-dissipation theorem \cite{Bertini2021May} is 
\be
-\frac{\Im\chi^R(\omega)}{\omega} = \frac{1 - e^{-\beta \omega}}{\omega} \Im \Ff[\Pi^R](\omega)  
\label{f-d equation}
\ee
where 
\be
\Pi^R(t) \equiv \Theta(t)\frac{\ex{\hat{I}(t)\hat{I}(0)}}{\i L} 
\label{pi^R definition}
\ee
Note that the superscript $R$ denotes the retarded function and that's why such functions come with $\Theta$-functions in their definitions.

For small $\omega$, the Taylor of the right-hand side reduces to $-\beta  \Im \Ff[\Pi^R](0) +\Oo(\omega)$, where the leading order term is the DC conductivity $\left( \sigma^{\text{DC}} = \lim_{\omega \to 0} \left( - \Im \chi^R(\omega)/\omega \right)\right)$ 
\begin{equation}
   \sigma^{\text{DC}} = \frac{\beta}{L}\, \Re \nint[0][\infty]{t} \ex{\hat{I}(t)\hat{I}(0)}=   \frac{\beta}{L}\, \Re \nint[-\infty][0]{t} \ex{\hat{I}(0)\hat{I}(t)}
    \label{dc conductivity formula}
\end{equation}

Therefore, we have obtained the expressions for dynamical conductivity (Eq. \eqref{dynamical conductivity formula}) and DC conductivity (Eq. \eqref{dc conductivity formula}) where we see that the central object for all our calculations is the current-current correlation $\ex{\hat{I}(0)\hat{I}(t)}$. We define the following integral that will play the central role in all our calculations (note the time-ordering in the integrand): 
   \begin{equation}
    \Ww_{\kappa}(t) \equiv  \nint[-\infty][t]{\tau} \frac{2\, \ex{\hat{I}(0) \hat{I}(\tau)}}{\kappa N L} 
    \label{main integral}
    \end{equation}
We evaluate this integral for three different cases, namely $\kappa \in \{1/2,1,2\}$ later in this work. To summarize, we have the following quantities when expressed in terms of this integral:
\begin{equation}
   \sigma_{\kappa}(t) =  \kappa N \Im \Ww_{\kappa}(t) 
    \label{conductivity and conductance formula}
\end{equation}
with initial condition for $\sigma(t=0)$ given in Eq. \eqref{initial condition for sigma equation}, and
\begin{equation}
    \sigma_{\kappa}^{\text{DC}} = \beta \kappa N \Re \Ww_{\kappa}(0)/2 
    \label{sigma DC in terms of W}
\end{equation}
where $\kappa$ is used as an index for the three cases we are interested in this work, namely $\kappa = \{1/2, 1, 2\}$. 

\section{Current-Current Correlations}
\label{current correlations section}

After developing the formalism in detail and for general situations, we realize that the central object for all our calculations is the current-current correlations $\ex{\hat{I}(0)\hat{I}(t)}$ that appears in the main integral in Eq. \eqref{main integral} in terms of which we can express dynamical and DC conductivities as well as conductance as in Eqs. \eqref{conductivity and conductance formula} and \eqref{sigma DC in terms of W}. Now we evaluate the current-current correlation in this section. Without loss of generality, we consider $t>0$ and therefore only the forward contour in the Keldysh plane contributes in Fig. \ref{keldysh plane}. Therefore, we are interested in deforming only the forward contour, which implies setting $\vec{\et}$ to zero for our uniformly coupled equilibrium case in Eq. \eqref{equilibrium effective action in terms of lambda and eta}. This simplifies to 
    \begin{equation}
    \begin{aligned}
        \frac{S_I}{|D|^2}  &= S_I |_{\vec{\ll}= 0} \\
        &+ \frac{1}{2} \sum\limits_{n>m} \left[ \ll_n \llbar_m F(\tau_n - \tau_m)  + \llbar_n \ll_m F(\tau_n - \tau_m)   \right]  \\
        &- \i \sum_m \left[  \llbar_m + \ll_m  \right] \left( \nint[-\infty][0]{t}  Y(t) \right) +\frac{1}{2} \sum\limits_{n} \ll_n \llbar_n  X(0)
    \end{aligned}
    \label{forward contour deformation 1}
\end{equation}
where the indices $\{n,m\}$ can take values over all the contour-deformations. Below, we will see that these values generally range up till $2$ for all intents and purposes of our work (also see Appendix \ref{app. current-current correlation calculation}) and it's easily generalizable to higher order moments if required in future works. 

The interacting and free parts of the action appears in the Keldysh partition function as in Eq. \eqref{disorder averaged partition function} where the free part $S_0$ does not contain any couplings, therefore $S_0$ does not contain any deformations in terms of $\{\vec{\ll}, \vec{\et}\}$ where recall that $\vec{\ll} = \left( \{\ll_n \}, \{\llbar_m \} \right)$ and $\vec{\et} = \left( \{\et_n \}, \{\etbar_m \} \right)$, these only appear in the interacting part of the action $S_I$ as in Eq. \eqref{equilibrium effective action in terms of lambda and eta}. Therefore, the total action $S = S_0 + S_I$ will lead to expressions such as $\partial_{\ll_n} S = \partial_{\ll_n} S_I$. We also know that the total Keldysh partition function becomes unity in the absence of any contour deformation, namely 
\be \left.\Zz\right|_{\{\vec{\ll}, \vec{\et}\} = 0} = 1 . \label{keldysh partition function equals unity}\ee

We start by evaluating the expectation values of energy that plays a role, e.g., in the initial condition for $\sigma(t=0)$ in Eq. \eqref{initial condition for sigma equation}. We already have the expression in Eq. \eqref{expectation of H} where we deformed the forward contour just like what we are interested here because we have assumed $t>0$. So we have (using Eq. \eqref{forward contour deformation 1}) 
\begin{equation}
	\begin{aligned}
	\ex{\Hh_{\rightarrow}(t)} &= \i\p_{\ll_1} \Zz \vert_{\vec{\ll} = 0} =\i\p_{\ll_1} \int \mathcal{D}\Gg \mathcal{D}\Sigma e^{-\Nn L S(\ll)[\Gg, \Sigma]}\vert_{\vec{\ll} = 0}\\
	&= \i\int \mathcal{D}\Gg \mathcal{D}\Sigma \left(-\Nn L  [\p_{\ll_1} S_I(\ll)]\vert_{\vec{\ll} = 0}\right) e^{-\Nn L S(0)[\Gg, \Sigma]} \\
	&= - \Nn L |D|^2 \nint[-\infty][0]{t} Y(t)
	\end{aligned}
  \label{Etrans}
\end{equation}
where we used Eq. \eqref{keldysh partition function equals unity}. This can be used in the definition for $\hat{\Hh}_{\text{trans}}$ in Eq. \eqref{H dot and H trans defined} to get the expectation values of $\hat{\Hh}_{\text{trans}}$ that gets plugged into, e.g., Eq. \eqref{initial condition for sigma equation}.

Similarly, we can extract the average of product of transport Hamiltonian using Keldysh formalism. As an example, we have
\be \ex{\Hh_{\rightarrow}(0)\Hh_{\rightarrow}^\dag(t)} = -\p_{\ll_1}\p_{\bar{\ll}_2}\Zz \qquad (t>0). \label{expectation of H times H}\ee 
The time ordering is important here: it means that we cannot extract out expectation values such as $\ex{\Hh_{\rightarrow}^\dag(t)\Hh_{\rightarrow}(0)}$ using this method if $t>0$. Naively, one might assume that this is just $-\p_{\llbar_2}\p_{\ll_1} \Zz$. However, this would imply that the two operators commute because the derivatives commute, but this is not always the case as in general $[\Hh_{\rightarrow}(0), \Hh_{\rightarrow}^\dag(t)]\neq 0 $. Therefore, we have focused on our choice for $\ll_1$ and $\ll_2$ to maintain the time ordering since we are interested in objects of type $\ex{\hat{A}(t_1) \hat{B}(t_2)}$ where $t_1<t_2$ for any operators $\hat{A}$ and $\hat{B}$. We chose $t_1 = 0$ and $t_2 = t>0$ and identify $\ll_1$ with $t_1 = 0$ while $\ll_2$ with $t_2 = t$ to ensure $\ll_2 > \ll_1$.

We are interested in the total current-current correlation, where we use the definition of the total current from Eq. \eqref{currentOp} ($\hat{\Hh}_{\rightarrow}$ is defined in Eq. \eqref{H rightarrow defined}) to get (recall $t>0$)
\begin{widetext}
    \begin{equation}
	\begin{aligned}
		\ex{\hat{I}(0)\hat{I}(t)} &= \frac{1}{4} \ex{(\i r \Hh_{\rightarrow}(0)  - \i r \Hh_{\rightarrow}^\dag(0))(\i r \Hh_{\rightarrow}(t)  - \i r  \Hh_{\rightarrow}^\dag(t))}\\
		&=- \frac{r^2}{4} \ex{[\Hh_{\rightarrow}(0)  -  \Hh_{\rightarrow}^\dag(0)][\Hh_{\rightarrow}(t)  - \Hh_{\rightarrow}^\dag(t)]}\\
		&= \frac{r^2}{4} \left[ \left( \ex{\Hh_{\rightarrow}(0)  \Hh_{\rightarrow}^\dag(t)}    + \ex{\Hh_{\rightarrow}^\dag(0) \Hh_{\rightarrow}(t)}  \right) -  \ex{\Hh_{\rightarrow}^\dag(0) \Hh_{\rightarrow}^\dag(t)}- \ex{\Hh_{\rightarrow}(0) \Hh_{\rightarrow}(t)} \right]
	\end{aligned}
\end{equation}
\end{widetext}
where we explicitly extracted the transport couplings out of $\Hh_{\rightarrow}$ and $\Hh_{\rightarrow}^\dagger$ in the last equality. Then using Eq. \eqref{expectation of H times H}, we can evaluate this current-current correlation In terms of the partition function as
\begin{equation}
\ex{\hat{I}(0)\hat{I}(t)} = - \frac{r^2}{4} ( \p_{\ll_1}\p_{\bar{\ll}_2}+ \p_{\bar{\ll}_1}\p_{\ll_2}  - \p_{\bar{\ll}_1}\p_{\bar{\ll}_2} - \p_{\ll_1}\p_{\ll_2})\Zz|_{\vec{\ll} = 0}
	\label{current correlation in terms of derivatives}
\end{equation}
We evaluate each of the terms on the right-hand side in Appendix \ref{app. current-current correlation calculation} to which we refer the readers for detailed calculations. The final general result for $r$-body hopping in Eq. \eqref{model equation} ($r=\kappa q$ for our case) is given by (see Appendix \ref{app. current-current correlation calculation} for details)
\begin{equation}
     \ex{\hat{I}(0)\hat{I}(t)} = + \frac{r^2}{4} NL |D|^2 F(t)
     \label{current correlation in terms of F main text}
\end{equation}
which specializes to our case to (again, see Appendix \ref{app. current-current correlation calculation} for details)
\begin{equation}
    \ex{\hat{I}(0)\hat{I}(t)}_{\kappa} = +\frac{\kappa}{8} NL |\Dd|^2e^{\kappa g^+(t)} 
    \label{current correlation for our case main text}
\end{equation}
where we have introduced a label $\kappa$ in the subscript to denote the three cases we are interested in, namely $\kappa = \{1/2, 1, 2\}$. Recall that the charge density scales as $\Qq = \Oo(1/\sqrt{q})$ and  we have defined
\begin{equation}
 |\Dd|^2 \equiv   2 |D|^2 e^{-2 \kappa q \Qq^2}
 \label{curly D definition}
\end{equation}
in the same spirit as we defined $\Jj_q$ and $\Kk_{\kappa q}$ in Eq. \eqref{curly J definition} where $\Dd = \Oo(q^0)$ in the large-$q$ limit. There is another advantage in defining $\Dd$ with a factor of $2$, namely that the mapping between the uniformly coupled chain at equilibrium in Eq. \eqref{model equation} to the dot in Eq. \eqref{model equivalent equation} which is given by $J^2 \to J_q^2$ and $2 |D|^2 \to K_{\kappa q}^2$ (see below Eq. \eqref{model equivalent equation}) becomes $\{ \Jj^2 \to \Jj_q^2, |\Dd|^2 \to \Kk_{\kappa q}^2 \}$. Plugging this in the main integral in Eq. \eqref{main integral}, we get
\begin{equation}
    \mathcal{W}_{\kappa} (t)=  \int_{-\infty}^t d\tau \frac{|\Dd|^2}{4} e^{\kappa g^+(\tau)}
    \label{main integral for our case}
\end{equation}
as our main integral in focus that is connected to the dynamical conductivity and DC conductivity via Eqs. \eqref{conductivity and conductance formula} and \eqref{sigma DC in terms of W}.

\section{Conductivity}
\label{conductivity section}

We are interested in three cases, namely $\kappa = \{1/2, 1, 2 \}$ whose exact analytical solution at leading order in $1/q$ is presented in Section \ref{analytical solution subsection}. Therefore, we know the exact expression for $g^+(t)$ for these three cases, thereby implying that we know the current-current correlations using Eq. \eqref{current correlation for our case main text} in an exact closed analytical form. Having evaluated the central object for all intents and purposes, we can use this to evaluate the main integral in Eq. \eqref{main integral} (simplified to Eq. \eqref{main integral for our case} for our case) related to the dynamical and DC conductivities in Eqs. \eqref{conductivity and conductance formula} and \eqref{sigma DC in terms of W}, respectively. We reproduce them here for convenience:
\begin{equation}
    \begin{aligned}
        \Ww_{\kappa}(t) &\equiv  \nint[-\infty][t]{\tau} \frac{2\, \ex{\hat{I}(0) \hat{I}(\tau)}}{\kappa N L} =  \int_{-\infty}^t d\tau \frac{|\Dd|^2}{4} e^{\kappa g^+(\tau)} \\
\sigma_{\kappa}^{\text{DC}} &= \beta \kappa N \Re \Ww_{\kappa}(0)/2 
    \end{aligned}
\end{equation}
where subscript $\kappa$ is a label for the three cases of $\kappa$ we are interested in. Here the function $g_+$ is the solution to the differential equation
\begin{equation}
    \ddot{g}^+(t) = -2|\Dd|^2 e^{\kappa g^+(t)}-2\Jj^2 e^{g^+(t)}, \label{kappaDe}
\end{equation}
 where $\Dd$ is the hopping strength and $\Jj$ is the on-site SYK strength where we have used the mapping between a uniformly coupled chain at equilibrium (Eq. \eqref{model equation}) and the dot (Eq. \eqref{model equivalent equation}), namely $\{ \Jj^2 \leftrightarrow \Jj_q^2, |\Dd|^2 \leftrightarrow \Kk_{\kappa q}^2 \}$ in the equations in Section \ref{analytical solution subsection} where we provided the exact analytical solutions to the three cases at leading order in $1/q$.

\subsection{Integrals of interest}\label{IntDetails}

We will see how these integrals all have solutions in terms of the following functions
\begin{equation}
\begin{aligned}
    \Upsilon_{s,\gamma}(\theta) &\equiv \i \left[\text{Li}_s(-e^{\theta+\i\gamma})-\text{Li}_s(- e^{\theta-\i\gamma})\right]\\
    &= -2 \text{Li}_{s-1}(-e^{\theta})\gamma + \Oo(\gamma^3), 
    \end{aligned}
    \label{PolInts}
\end{equation}
where $\text{Li}_0(z) \equiv z/(1-z)$. The polylogarithm function is defined by $\text{Li}_{1}(z) = -\ln(1-z)$ and the recursion relation $\p_{\theta}\text{Li}_{s+1}(a e^{\theta}) = \text{Li}_{s}(a e^{\theta})$. It also has the power series representation
\begin{equation}
    \text{Li}_{s}(z) = \sum_{k=1}^{\infty} \frac{z^k}{k^s}
\end{equation}
for $|z|<1$. Note that $\text{Li}_{s}(0)=0$, as such $\Upsilon_{s,\gamma}(-\infty) = 0$. Further, for $s\ge 0$ as the real part of $z$ goes to $\infty$, $\text{Li}_{s}(e^{z}) \to -z^s/s!$. As such, we have  
\begin{equation}
    \Upsilon_{s,\gamma}(\theta) \xrightarrow[]{s\geq 0, z \to \infty}  \i \frac{(\theta-\i\gamma)^s-(\theta+\i\gamma)^s}{s!}
\end{equation}

We are interested in integrals of $\Upsilon_{s}(\theta)$ which are simply given by
\begin{equation}
    \nint{\theta} \Upsilon_{s,\gamma}(\theta) = \Upsilon_{s+1,\gamma}(\theta)
    \label{integration identity for polylogarithm function}
\end{equation}
where we have
 \begin{equation}
 \begin{aligned}
     \Upsilon_{0,\gamma}(\theta) &= \frac{\tan\gamma}{1+\sqrt{1+\tan^2\gamma} \cosh\theta} \\
     \Upsilon_{1,\gamma}(\theta) &= \gamma+2 \tan^{-1}\left[\tan(\gamma/2)\tanh(\theta/2) \right] \\
     &= [1+\tanh(\theta/2)]\gamma + \Oo(\gamma^3)
     \end{aligned}
     \label{special functions defined}
 \end{equation}

\subsection{\texorpdfstring{$\kappa = 1/2$}{}: Pseudo-kinetic case}
\label{k=1/2 conductivity subsection}

The solution for $\kappa = 1/2$ is given in Eq. \eqref{k=1/2 solution}. Using the mapping, namely $\{ \Jj^2 \leftrightarrow \Jj_q^2, |\Dd|^2 \leftrightarrow \Kk_{\kappa q}^2 \}$, we reproduce the differential equation for $g^+(t)$ (Eq. \eqref{k=1/2 differential equation}) along with its solution for convenience
\begin{equation}
    \begin{aligned}
        \ddot{g}^{+}(t)&=-2 \Jj^2 e^{g^{+}(t)}-2|\Dd|^2 e^{g^{+}(t)/2} \\
        e^{g^{+}(t) / 2}&=\frac{1}{\left(\beta |\Dd|\right)^2} \frac{(\pi v)^2}{1+\sqrt{A^2+1} \cos (\pi v(1 / 2-\imath t / \beta))}
    \end{aligned}
    \label{k=1/2 equation and solution in conductivity section}
\end{equation}
where $A \equiv \frac{\pi v \beta \mathcal{J}}{\left(\beta |\Dd|\right)^2}$ (see Eq. \eqref{A definition}). The closure relation is (Eq. \eqref{k=1/2 closure relation})
\begin{equation}
\pi v=\sqrt{\left(\beta \mathcal{J}\right)^2+\left(\frac{\left(\beta |\Dd| \right)^2}{\pi v}\right)^2} \cos (\pi v / 2)+\frac{\left(\beta |\Dd| \right)^2}{\pi v}
\label{full closure relation k=1/2 in conductivity subsection}
\end{equation}
which at low-temperatures becomes (see Re. \cite{Louw2023Dec} for further details about this $\kappa = 1/2$ case in thermodynamic and dynamic sense)
\begin{equation}
    v \xrightarrow[]{\text{low-temperature}} 2 - 2 \alpha_{1/2} T + \Oo(T^2)
    \label{k=1/2 closure relation at low temperature}
\end{equation}
where
\begin{equation}
    \alpha_{1/2} \equiv \frac{2}{|\Dd|}\sqrt{2 + |\Jj/\Dd|^2}
\end{equation}
The main integral in Eq. \eqref{main integral for our case} takes the form
\begin{equation}
    \Ww_{\kappa=1/2}(t) =  \int_{-\infty}^t d\tau \frac{|\Dd|^2}{4} e^{ g^+(\tau)/2} 
\end{equation}
Then we can re-cast the integrand in terms of the polylogarithm function as
\begin{equation}
\frac{|\Dd|^2}{4} e^{g^{+}(t)/2} = \frac{1}{4}\frac{d\theta}{d t}\frac{\pi v T}{\tan\gamma_{1/2}}\Upsilon_{0,\gamma_{1/2}}(\theta), 
\end{equation}
where
\begin{equation}
    \tan\gamma_{1/2} \equiv \pi v T \Jj |\Dd|^{-2},\quad \theta \equiv \pi v T t + \i \pi v/2
    \label{gamma and theta defined for k=1/2}
\end{equation}
and $\Upsilon_{0,\gamma_{1/2}}$ is given in Eq. \eqref{special functions defined}. This relation for the integrand can be checked by plugging in the expressions explicitly and using the identity $\cosh(\i x) = \cos(x)$. Therefore, the integral becomes
\begin{equation}
    \begin{aligned}
       \Ww_{1/2}(t) &=  \int_{-\infty}^t d\tau \frac{|\Dd|^2}{4} e^{ g^+(\tau)/2}  \\
       &= \int_{-\infty}^t d\tau \frac{1}{4}\frac{d\theta}{d \tau}\frac{\pi v T}{\tan\gamma_{1/2}}\Upsilon_{0,\gamma_{1/2}}(\theta)\\
       &= \frac{1}{4}\frac{\pi v T}{\tan\gamma_{1/2}} \int d\theta \Upsilon_{0,\gamma_{1/2}}(\theta)\\
       &= \frac{1}{4}\frac{\pi v T}{\tan\gamma_{1/2}} \Upsilon_{1,\gamma_{1/2}}(\theta)
    \end{aligned}
\end{equation}
where we used the integration identity in the last step as given in Eq. \eqref{integration identity for polylogarithm function}. The expression for $\Upsilon_{1,\gamma_{1/2}}(\theta)$ is given in Eq. \eqref{special functions defined} where we use the small $\gamma_{1/2}$ expression since we are interested in the low-temperature limit. Thus, using the limit $\lim_{x \to 0} x/\tan(x) = 1$, we get
\begin{equation}
    \begin{aligned}
\Ww_{1/2}(t) &= \frac{1}{4} \pi v T \left( 1 + \tanh(\theta/2)\right)         \\
&= \frac{1}{4} \pi v T \left( 1 + \tanh(\pi v T t/2 + \i \pi v/4)\right)   
    \end{aligned}
\end{equation}
This leads us to the expression for DC conductivity where we use the identity $\tanh(\i x) = \i \tan(x)$
\begin{equation}
\sigma_{\kappa = 1/2}^{\text{DC}} = \beta  N \Re \Ww_{\kappa}(0)/4 = \frac{N \pi v}{16}
\label{conductivity across all temp for k=1/2}
\end{equation}
which at low-temperature takes the form using Eq. \eqref{k=1/2 closure relation at low temperature}
\begin{equation}
    \sigma_{\kappa = 1/2}^{\text{DC}} \xrightarrow[]{\text{low-}T} \frac{N\pi}{8} \left(1 -  \alpha_{1/2} T\right)
    \label{sigma dc for k=1/2}
\end{equation}
which takes the following universal value independent of coupling constants $\Jj$ and $|\Dd|$ at zero-temperature:
\begin{equation}
    \sigma^{\text{DC}}_{\kappa =1/2}(T=0) = \frac{N \pi}{8} \quad (\text{maximal conductivity})
 \label{universal k=1/2 conductivity}
\end{equation}
This is the universal residual conductivity which is also the maximal conductivity across all temperatures. This matches the residual (and maximal) conductivity for $\kappa =1$ case (see Eq. \eqref{universal k=1 conductivity} below) when $\Jj = 0$ in $\kappa =1$ case, as well as Eq. \eqref{universal k=2 conductivity} for $\kappa = 2$ case. 

Finally, we invert the relation to leading order in temperature to get the resistivity
\begin{equation}
    \rho^{\text{DC}}_{\kappa = 1/2} = \frac{8}{N \pi} \left(1 + \alpha_{1/2} T\right) 
    \label{rho dc for k=1/2}
\end{equation}
which is linear-in-temperature as expected from a strange metal. This conducting phase is what we also get in $\kappa=1$ case (see Eq. \eqref{rho dc for k=1} below).

\subsection{\texorpdfstring{$\kappa = 1$}{}}
\label{k=1 conductivity subsection}
The exact solution for $g^+(t)$ is given in Eq. \eqref{k=1 solution}. We can resort to the formalism of polylogarithm function as we did in $\kappa =1/2$ as well as $\kappa =2$ (see below) but we can also directly deal with the explicit solution for $g^+(t)$ in Eq. \eqref{k=1 solution} and the differential equation for $g^+(t)$ in Eq. \eqref{k=1 differential equation} (again, after using the mapping from the dot to the chain) which we reproduce here for convenience
\begin{equation}
    \begin{aligned}
         \ddot{g}^+(t) &= -2 \left( |\Dd|^2 +\Jj^2\right) e^{g^+(t)} \\
     e^{g_+(t)} &= \frac{(\pi v)^2}{\beta^2(\Jj^2 + |\Dd|^2) \cos^2(\pi v(1/2 - \i t/\beta))} 
    \end{aligned}
    \label{k=1 equation and solution in conductivity section}
\end{equation}
with the closure relation (Eq. \eqref{k=1 closure relation}) 
\begin{equation}
    (\pi v)^2= \beta^2 \left(  \Jj^2  +  |\Dd|^2 \right) \cos^2 (\pi v / 2),
    \label{full closure relation k=1 in conductivity subsection}
\end{equation}
which at low-temperature becomes
\begin{equation}
    v \xrightarrow[]{\text{low-temperature}} 1 - \alpha_1 T + \Oo(T^2)
    \label{k=1 low temp expansion of v}
\end{equation}
where
\begin{equation}
    \alpha_1 \equiv \frac{2}{|\Dd|}\frac{1}{\sqrt{1+|\Jj/\Dd|^2}}.
\end{equation}

Since we are interested in evaluating the main integral as in Eq. \eqref{main integral for our case}, the differential equation in Eq. \eqref{k=1 equation and solution in conductivity section} can be rewritten to match the integrand of the main integral as
\begin{equation}
    \frac{|\Dd|^2}{4} e^{g^+(t)} = \frac{-|\Dd|^2}{8 (|\Dd|^2 + \Jj^2)} \ddot{g}^+(t)
\end{equation}
which can be integrated to get for the main integral
\begin{equation}
        \Ww_{\kappa = 1}(t) =  \frac{|\Dd|^2}{4} \nint[-\infty][t]{t} e^{g_+(t)}  =  \frac{\dot{g}_+(-\infty) -\dot{g}_+(t)}{8|\Jj/\Dd|^2 + 8}
    \end{equation}
Using the solution for $g^+(t)$ from Eq. \eqref{k=1 equation and solution in conductivity section}, we get
\begin{equation}
    \dot{g}^+(t) = -2 \i \pi  v T \tan \left(\pi  v (1/2 -\i t/\beta)\right)
\end{equation}
where
\begin{equation}
    \lim\limits_{t\to -\infty} \dot{g}^+(t) = 2 \pi v T \in \mathbb{R}
\end{equation}
Therefore, we have for the main integral
\begin{equation}
    \Ww_{1}(t) = \frac{\pi v T}{4|\Jj/\Dd|^2 + 4} \left(1 +  \i \tan\left[\pi  v (1/2 -\i t/\beta)\right] \right)
\end{equation}
Hence we have for the DC conductivity
\begin{equation}
    \sigma^{\text{DC}}_{\kappa =1} = \beta N \Re \Ww_1(0)/2 = \frac{N \pi v}{8(|\Jj/\Dd|^2 + 1)}
    \label{conductivity across all temp for k=1}
\end{equation}
where we plug the low-temperature expansion of $v$ in Eq. \eqref{k=1 low temp expansion of v} to get
\begin{equation}
    \sigma^{\text{DC}}_{\kappa =1} \xrightarrow[]{\text{low-}T} \frac{N \pi }{8(|\Jj/\Dd|^2 + 1)} \left(1 - \alpha_1 T\right)
    \label{sigma dc for k=1}
\end{equation}
We find a universal residual DC conductivity at $T=0$ which is also the maximal conductivity across all temperatures for $\Jj = 0$ as
\begin{equation}
    \left.\sigma^{\text{DC}}_{\kappa =1}(T=0)\right|_{\Jj = 0} = \frac{N \pi}{8} \quad (\text{maximal conductivity})
    \label{universal k=1 conductivity}
\end{equation}
which is the same as we found as the universal residual DC conductivity which was also the maximal conductivity across all temperatures for $\kappa = 1/2$ case in Eq. \eqref{universal k=1/2 conductivity} as well as Eq. \eqref{universal k=2 conductivity} for $\kappa = 2$ case. Note that for $\kappa = 1/2$, the residual (and maximal) DC conductivity is independent of coupling constants while $\kappa = 1$ case depends on the coupling constants and both matches when $\Jj=0$ in $\kappa = 1$ case. 

Finally, we invert at low-temperature (to order $\Oo(T)$) to get the resistivity
\begin{equation}
    \rho^{\text{DC}}_{\kappa = 1} = \frac{8(|\Jj/\Dd|^2 + 1) }{N \pi} \left(1 + \alpha_1 T\right) 
    \label{rho dc for k=1}
\end{equation}
which is linear-in-temperature as expected from a strange metal. This is the same as we got for the $\kappa =1/2$ case in Eq. \eqref{rho dc for k=1/2}.


\subsection{\texorpdfstring{$\kappa = 2$}{}: Insulator case}
\label{k=2 conductivity subsection}

Just like the $\kappa =1/2$ case in Section \ref{k=1/2 conductivity subsection}, the differential equation and solution for $g^+(t)$ (Eqs. \eqref{k=2 differential equation} and \eqref{k=2 solution} after using the chain-dot mapping as before) is reproduced here as
\begin{equation}
    \begin{aligned}
     \ddot{g}^{+}(t) &= -2 |\Dd|^2 e^{2 g^{+}(t)} - 2 \Jj^2 e^{g^{+}(t) }\\
     e^{g^{+}(t) }&=\frac{1}{2\left( \beta \mathcal{J}\right)^2} \frac{(\pi v)^2}{1+\sqrt{B^2+1} \cos (\pi v(1 / 2-\imath t / \beta))}
    \end{aligned}
    \label{k=2 equation and solution in conductivity section}
\end{equation} 
where $B \equiv \frac{ \pi v \beta |\Dd|}{\sqrt{2} \left( \beta \mathcal{J} \right)^2} = \frac{ \pi v T |\Dd|}{\sqrt{2}  \mathcal{J}^2} $. The closure relation is (Eq. \eqref{k=2 closure relation})
\begin{equation}
\pi v=\sqrt{2 \left( \beta |\Dd|\right)^2+\left(\frac{2\left( \beta \mathcal{J}\right)^2}{\pi v}\right)^2} \cos (\pi v / 2)+\frac{2 \left( \beta \mathcal{J}\right)^2}{\pi v}
\label{k=2 closure relation in conductivity subsection}
\end{equation}
which at low-temperature in leading order in temperature becomes
\begin{equation}
    v \xrightarrow[]{\text{low-temperature}} 2 - 2\, \alpha_2 T+\Oo(T^2)
\end{equation}
where 
\begin{equation}
    \alpha_2 \equiv \frac{\sqrt{2}}{ \Jj}\sqrt{2+|\Dd/\Jj|^2}.
\end{equation}
The main integral (Eq. \eqref{main integral for our case}) is given by
\begin{equation}
    \Ww_{\kappa=2}(t) =   \int_{-\infty}^t d\tau \frac{|\Dd|^2}{4} e^{2 g^+(\tau)} .
\end{equation}
The integrand can be re-written using the differential equation in Eq. \eqref{k=2 equation and solution in conductivity section} as
\begin{equation}
    \frac{|\Dd|^2}{4} e^{2g^+(\tau )} = -\frac{\ddot{g}(\tau)}{8} -  \frac{\Jj^2}{4} e^{g^+(\tau)}.
\end{equation}
The second expression can be expressed by using the solution for $g^+(t)$ from Eq. \eqref{k=2 equation and solution in conductivity section}
\begin{equation}
    \frac{\Jj^2}{4} e^{g^+(\tau)} = \frac{1}{8}\frac{\pi v T}{\tan \gamma_{2}} \frac{d \theta}{d t}   \Upsilon_{0,\gamma_{2}}(\theta)
\end{equation}
where
\begin{equation}
    \tan\gamma_{2} \equiv \pi v T \frac{|\Dd|}{\sqrt{2} \Jj^{2}} ,\quad \theta \equiv \pi v T t + \i \pi v/2
\end{equation}
where $\theta$ is the same as defined in $\kappa = 1/2$ case (Eq. \eqref{gamma and theta defined for k=1/2}) and the expression for $\Upsilon_{0,\gamma_{2}}(\theta)$ is given in Eq. \eqref{special functions defined}. By plugging the explicit expressions, this can be directly verified. Note that we use identity $\cos(\i x) = \cos(x)$. Therefore, the integrand becomes
\begin{equation}
     \frac{|\Dd|^2}{4} e^{2g^+(\tau )} = -\frac{\ddot{g}(\tau)}{8} - \frac{1}{8}\frac{\pi v T}{\tan \gamma_{2}} \frac{d \theta}{d t}   \Upsilon_{0,\gamma_{2}}(\theta).
\end{equation}
Thus the integral becomes
\begin{equation}
    \Ww_2 (t) = \frac{\dot{g}(-\infty)-\dot{g}(t)}{8} - \frac{1}{8} \frac{\pi v T}{\tan \gamma_2} \Upsilon_{1,\gamma_{2}}(\theta)
\end{equation}

Now we simplify $\dot{g}(t)$ which is obtained from Eq. \eqref{k=1 equation and solution in conductivity section} as (recall $\i \sin(x) = \sinh(x)$ and $\cosh(- \i x) = \cos(x)$)
\begin{equation}
   \dot{g}(t)= -\pi  v T \frac{\sinh \left(\i \pi \frac{v}{2} + \pi v T t\right)}{\frac{1}{\sqrt{1 +  \left( \frac{ \pi   v T |\Dd|}{ \sqrt{2} \Jj^2}\right)^2} }+ \cosh \left( \i \pi \frac{v}{2} + \pi v T t \right)}
\end{equation}
which can be re-written as
\begin{equation}
\begin{aligned}
    \dot{g}(t) &= - \pi v T \frac{\sinh \theta}{\frac{1}{\sqrt{1 + \tan^2 \gamma_2}} + \cosh \theta} \\
    &= - \pi v T \frac{\sinh \theta}{\cos \gamma_2 + \cosh \theta} 
    \end{aligned}
\end{equation}
So we have
\begin{equation}
    \dot{g}(-\infty) = +\pi v T
\end{equation}
Therefore
\begin{equation}
\begin{aligned}
    \Ww_2(t=0) = \frac{\pi v T}{8} \left[  1+\right. &\left. \frac{\sinh \left(\i \pi \frac{v}{2} \right)}{\frac{1}{\sqrt{1 +  \left( \frac{ \pi   v T |\Dd|}{ \sqrt{2} \Jj^2}\right)^2} }+ \cosh \left( \i \pi \frac{v}{2}  \right)}  \right.\\
    &\left. - \frac{1}{\tan \gamma_2} \Upsilon_{1,\gamma_{2}}(\i \pi v/2) \right]
\end{aligned}
\end{equation}
where we use the definition of $\Upsilon_{1,\gamma_{2}}(\theta) $ from Eq. \eqref{special functions defined} along with the identities $\sinh(\i x) = \i \sin x$, $\cosh(\i x) = \cos x$ and $\tanh(\i x) = \tan x$ to get
\begin{equation}
\begin{aligned}
    \Ww_2(t=0) = \frac{\pi v T}{8} \left[  1+\right. &\left. \frac{\i \sin \left( \pi \frac{v}{2} \right)}{\frac{1}{\sqrt{1 +  \left( \frac{ \pi   v T |\Dd|}{ \sqrt{2} \Jj^2}\right)^2} }+ \cos \left(  \pi \frac{v}{2}  \right)}  \right.\\
    &\left.  - \frac{\gamma_2}{\tan \gamma_2} (1 + \i \tan(\pi v /4)) \right]
\end{aligned}
\end{equation}
So we have for the DC conductivity the following: 
\begin{equation}
    \begin{aligned}
         \sigma^{\text{DC}}_{\kappa =2} =& \beta N \Re \Ww_2(0) = \frac{N \pi v }{8} \left( 1 - \frac{\gamma_2}{\tan \gamma_2}\right) \\
         =& \frac{N \pi v \gamma_2^2}{24} + \Oo(\gamma_2^4)
    \end{aligned}
    \label{sigma DC final result for k=2 for low temperature}
\end{equation}
where we used the expansion $1 - \frac{x}{\tan x} = \frac{x^2}{3} + \Oo(x^4)$ in the second line. We use the small $\gamma_2$ limit (low-temperature limit) to get $\tan \gamma_2 \approx \gamma_2 = \pi v T \frac{|\Dd|}{\sqrt{2} \Jj^{2}}$ and $v \approx 2 - 2\, \alpha_2 T+\Oo(T^2)$ to get in leading order in temperature, in this case the order being $\Oo(T^2)$, as
\begin{equation}
  \sigma^{\text{DC}}_{\kappa =2} =  \frac{N \pi^3 v^2 T^2 |\Dd|^2}{48 \Jj^2} + \Oo(T^4)=  \frac{N \pi^3  T^2 |\Dd|^2}{12 \Jj^4} + \Oo(T^3)
  \label{sigma DC for k=2 explicit in temperature}
\end{equation}
which corresponds to an insulator resistivity $\rho^{\text{DC}}_{\kappa = 2} \sim T^{-2}$. This also matches the claim found in footnote 18 of Ref. \cite{Cha2020Sep}. We remind the readers that the exact analytical solution to leading order in $1/q$ was presented in Section \ref{analytical solution subsection} where we noted that the solutions for $\kappa = 2$ case is obtained from $\kappa = 1/2$ case by a simple substitution of $\left. (g^+, \Kk_{q/2}^2, \Jj_q^2)\right|_{\kappa = 1/2} \to \left. 2 \left(g^+, \Jj_q^2, \Kk_{2q}^2 \right)\right|_{\kappa = 2}$. Despite this simple transformation to obtain the exact solution, the physics of $\kappa =1/2$ and $\kappa = 2$ cases differ drastically when it comes to transport properties as we showed now, namely that the former is a conductor while the latter is an insulator. Finally, we note from the first line in Eq. \eqref{sigma DC final result for k=2 for low temperature} that the maximal conductivity is given by
\begin{equation}
    \sigma^{\text{DC}}_{\kappa = 2}|_{\text{max}} = \frac{N \pi}{8 } \quad (\text{maximal conductivity})
    \label{universal k=2 conductivity}
\end{equation}
This matches with the maximal conductivity found in $\kappa = 1/2$ (Eq. \eqref{universal k=1/2 conductivity}) and $\kappa = 1$ (Eq. \eqref{universal k=1 conductivity}) cases. Note that the maximal conductivity matches with the residual conductivity (conductivity at $T=0$) in $\kappa =1/2$ and $\kappa = 1$ cases in addition to other constraints on the coupling constants, while here the maximal conductivity happens at a non-zero temperature. We now proceed to study and plot the various resistivities across all three cases and discuss the universal properties that emerge.

\subsection{Resistivity discussion for all temperatures}

\begin{figure}
    \centering
    \includegraphics[width=\columnwidth]{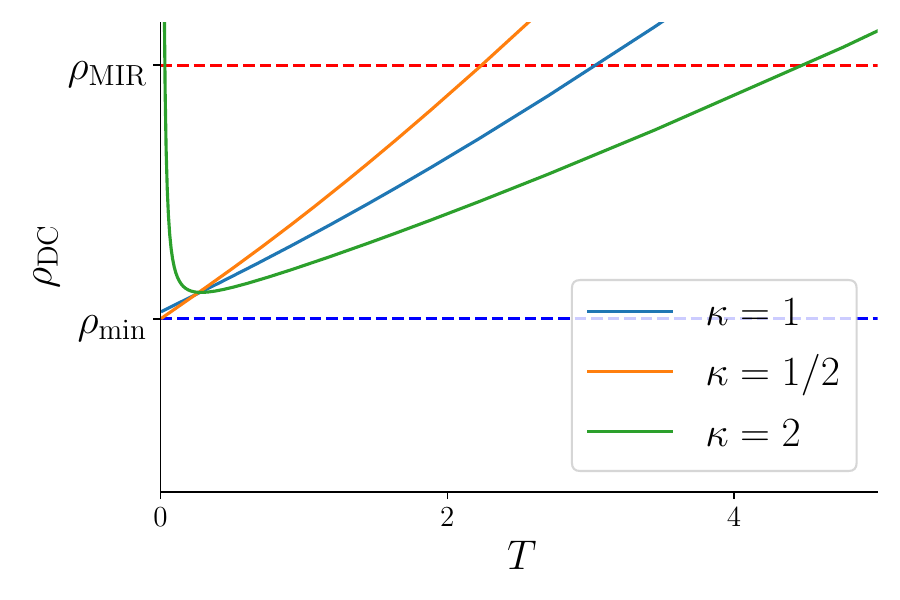}
    \caption{Resistivity divided by the MIR bound as a function of temperature for all three cases where $\Jj = 1$ and $|\Dd| = 5$. Note that $\kappa = 1/2$ and $\kappa =1 $ cases already have linear-in-temperature resistivity while $\kappa = 2$ case behaves as an insulator at low enough temperatures while it regains its linear-in-temperature resistivity at higher temperatures. We have a universal minimum resistance $\rho_{\text{min}} = 8/(N \pi )$ which is obtained for all values of $|\Dd|$ and $\Jj$ for the $\kappa = 1/2$ case, while it is obtained for particular values of coupling constants for $\kappa = 1$ and $\kappa = 2$ cases. Here we have shown the plot for only the particular values of $|\Dd| =5$ and $ \Jj = 1$ for which the two latter cases do not achieve the minimum resistivity.}
    \label{fig:Res}
\end{figure}

We make use of the results for DC conductivity expressions in Eqs. \eqref{conductivity across all temp for k=1/2}, \eqref{conductivity across all temp for k=1} and the first line of Eq. \eqref{sigma DC final result for k=2 for low temperature} for $\kappa = 1/2$, $\kappa = 1$ and $\kappa =2$ cases, respectively, where we used the closure relation for $v$ for all temperatures (Eqs. \eqref{full closure relation k=1/2 in conductivity subsection}, \eqref{full closure relation k=1 in conductivity subsection} and \eqref{k=2 closure relation in conductivity subsection}, respectively) instead of just the low-temperature expansion to plot the resistivities in Fig. \ref{fig:Res}. For all three cases our results are valid for \textit{all} temperature range including $T=0$, since we do not make use of the conformal solutions anywhere. The most noticeable of the three is the $\kappa = 2$ case which has a diverging resistivity as $T \to 0$. In this sense, it is an insulator. The $1/T^2$ divergence here is quite different from the exponential stretched exponential divergence $\ln(\rho) \sim \beta^{\alpha}$ in typical insulators \cite{Mott1969Apr} for some $\alpha$ depending on the insulator.  

The ratio $|\Dd|/\Jj$ does not change the qualitative behavior. However, the larger the value of $|\Dd|$, the lower the minimum resistivity. We find that when $|\Dd| \to \infty$, all minimum resistivities go to the universal value of $8/(N\pi )$.  For small enough ratios $|\Dd|/\Jj$, all resistivities are below the MIR bound 
 \be \rho_{\text{MIR}} = \frac{2\pi}{ N},\ee 
 for lattice spacing $a=1$ and flavor charge $q_e =1$ \cite{Werman2016Feb,Hartnoll2015Jan}
at low enough temperatures. As such, they all have a strange metal phase.  

This should be contrasted to the previous linear-in-$T$ resistivity studied for a $q=4$ SYK lattice with quadratic hopping $\kappa q = 2$ \cite{Song2017Nov}. Here the linear in $T$ behavior was given by 
\be \rho/\rho_{\text{MIR}} \sim  \frac{T}{D_2^2/J_4}\ee
only when
\be 1 \lll  \frac{T}{D_2^2/J_4} \lll  \frac{J_4^2}{D_2^2}.\ee
In other words, the behavior is only far above the MIR bound, therefore representing a bad metal instead of a strange metal.
In our case, when the resistivities take the universal minimum value of $8/(N\pi )$, then we have
\be \rho_{\text{min}}/\rho_{\text{MIR}} = 4/\pi^2 < 1.\ee
In other words, the resistivity is linear below the MIR bound, and hence we have a true strange metal. This is quite relevant because this has been discussed in the literature \cite{Chowdhury2022Sep, sachdev-essay} that bad metals are easily obtained at higher temperatures, but it is difficult to find a smooth crossover from Fermi liquid behavior to a strange metal that eventually goes above the MIR bound, signalling a bad metal behavior. Here we have obtained true strange metal behavior for all three cases. Moreover, for $\kappa =2$ case we found four interesting behaviors: (1) an insulating behavior ($\rho \sim T^{-2}$) at low temperatures, (2) this smoothly crossovers to quadratic in temperature ($\rho \sim (T-T_{\text{min}})^2$ where $T_{\text{min}}$ is the temperature where resistivity is minimum, see Fig. \ref{fig:Res}) around the minimum resistivity point, signalling a Fermi liquid type description (see Section \ref{FLSection} below for more details), (3) then a smooth crossover to a strange metal ($\rho \sim T$ but below the MIR bound) and (4) eventually becoming a bad metal ($\rho \sim T$ but above the MIR bound). This is summarized in Fig. \ref{fig:Res}. To emphasize the importance of the four stages of $\kappa = 2$ case, we cite a sentence from Ref. \cite{Chowdhury2022Sep} which states that, ``While bad-metals can arise at very high temperatures for rather simple reasons, the key puzzle is often related to their smooth evolution into a low-temperature regime without any characteristic crossovers that defies Fermi liquid behavior.''

\subsubsection{Low-temperature behavior of insulators}
\label{HolographicInsulatorSection}

Our insulator displays a $T^{-2}$ dependence. This is markedly different behavior from typical insulators at low temperatures. In typical insulators, the resistivity increases exponentially as temperature decreases, following an activated behavior of the form $\rho(T) \sim \exp\left(\Delta/T\right)$. This behavior arises due to the energy gap $\Delta$ in the electronic structure, which requires thermal excitation to promote carriers across the gap, leading to conduction.

This deviation from the usual behavior of standard insulators is noteworthy. Such a $T^{-2}$ dependence is characteristic of holographic insulators \cite{Horowitz2012Jul,Mefford2014Oct,Andrade2018Oct}, as observed in certain models rooted in the framework of gauge/gravity duality. In our study, the model considered is a Conformal Field Theory (CFT) at low temperatures, where the scale invariance and the lack of an intrinsic energy scale might be the reason for such a temperature dependence; however, we leave a further discussion of this for the outlook.

\subsubsection{Intermediate and high-temperature behavior} 
\label{FLSection}

As the temperature increases, our insulator undergoes a crossover from this unconventional $T^{-2}$ resistivity to a more conventional quadratic temperature dependence $\rho(T) \sim T^2$ near the minimum resistivity.

Such a $T^2$ behavior is typically associated with Fermi-liquid systems, where electron-electron scattering dominates. At even higher temperatures, the resistivity crossovers to a linear dependence on temperature:
\begin{equation}
    \rho(T) \sim T
\end{equation}

This linear-in-$T$ resistivity is indicative of a crossover to a strange metal regime, commonly seen in strongly correlated electron systems. An analogous crossover is also observed in SYK chains with quadratic hopping \cite{Song2017Nov}. In that case, the quadratic in $T$ resistivity indeed implied Fermi-liquid quasiparticles. Here, however, we have no quadratic hopping terms, rather we have diffusive $\Oo(q)$-particle hopping. As such, the question of whether Fermi-liquid-like quasiparticles exist in our model is also left for future study and discussed further in the outlook.

\subsection{Scaling arguments and universality}
\label{scaling arguments and universality}

As mentioned in the introduction and Section \ref{a note on scaling arguments}, the naive expectations from scaling arguments \cite{Shankar1994Jan} show that $\kappa =1/2$ and $\kappa = 2$ cases would have conducting and insulating behaviors at low temperatures, respectively. However, as mentioned above, the scaling arguments are not rigorous away from the critical points for SYK-like systems and therefore cannot be used as a substitute for rigorous mathematical calculations to derive the resistivity properties. Moreover, it cannot lead to any intuition for $\kappa = 1$ case as well where both the on-site and the hopping terms compete. After detailed mathematical calculations, we showed that indeed the naive expectations from scaling arguments hold for both $\kappa =1/2$ and $\kappa = 1$ cases. 

We argue that the scaling arguments for the aforementioned two cases that we happened to show rigorously by mathematical calculations can lead to an understanding of residual conductivity as observed in Eq. \eqref{universal k=1/2 conductivity} for $\kappa =1/2$ case. The hopping term dominates as the temperature $T \to 0$ and therefore there is no room for any parameter to control the conductivity at exactly $T=0$. This explains why the $T=0$ result in Eq. \eqref{universal k=1/2 conductivity} for $\kappa =1/2$ case is independent of all coupling constants. The same holds for the residual conductivity (Eq. \eqref{sigma DC for k=2 explicit in temperature}) for the$\kappa = 2$ case, which identically vanishes as $T \to 0$ for all values of coupling constants. Note that the temperature at which the maximal conductivity happens for $\kappa = 2$ case is non-zero and therefore differs from the temperature for residual conductivity, i.e. $T=0$. This differs from $\kappa = \{1/2,1\}$ cases where the maximal conductivities appear at $T=0$. Finally, since both terms compete for $\kappa =1$ case, there can be a non-trivial dependence of conductivity at $T=0$ on the ratio of coupling constants $\Jj/\Dd$ which is indeed the case as can be seen from Eq. \eqref{sigma dc for k=1}. Therefore, to obtain the maximal conductivity which happens at $T=0$ for $\kappa = 1$ case, we have to further enforce $\Jj = 0$.

\section{Discussion and Conclusion}
\label{conclusion section}

We developed an analytic functional-based approach to calculate thermal expectation values of any time sequence ordered correlations in a $G-\Sigma$ formalism. We started by deriving exact general relations corresponding to contour deformations in the Keldysh plane that is a consequence of studying the linear response theory by linearly perturbing the system. To illustrate the methodology developed in this work, we considered three SYK chains at equilibrium whose linear response were studied by switching on an external field. The three cases corresponded to having large-$q$ complex SYK dots with nearest-neighbor $r$-body hopping where $r = \kappa q$ and $\kappa$ can take three values: $\{1/2, 1, 2 \}$. We provided a mathematical mapping to map a uniformly coupled chain at equilibrium (Eq. \eqref{model equation}) to a single SYK dot (Eq. \eqref{model equivalent equation}) where we proceeded to provide exact analytical solutions to Green's functions for all the three aforementioned cases. Using the developed formalism for Keldysh contour deformations, we proceeded to derive the formula for conductivity, which required calculation of current-current correlations. We performed all calculations analytically and exact at leading order in $1/q$ where we obtained expressions for DC conductivity and accordingly the resistivity for all temperatures and specialized to low-temperature range to study the strange metal behavior. We found for $\kappa = \{1/2, 1\}$ cases a conducting phase where resistivity goes as linear-in-temperature at low temperatures, emblematic of strange metal behavior. For $\kappa =2 $ case, we found an insulating phase where resistivity diverges as a power law in temperature ($\rho \sim T^{-2}$) at low-temperatures. We also compared the resistivities across all three cases of $\kappa$ for all values of temperature against the MIR bound and showed that for certain temperature range, the resistivities in all the three cases are indeed below the MIR bound, thereby showing a true strange metal behavior (see Fig. \ref{fig:Res}). For the $\kappa = 2$ case, we found four interesting stages of smooth crossovers as temperature increases, from an insulating phase to a Fermi liquid behavior to a strange metal and eventually becoming a bad metal. This is quite relevant in light of what has been discussed in the literature \cite{Chowdhury2022Sep, sachdev-essay} about the difficulties of obtaining such a behavior. We have obtained this analytically in the thermodynamic limit, enabling further investigations as discussed below.

Remarkably, we found a universal maximal DC conductivity ($=N\pi/8$) for all three cases of $\kappa$ across all temperature ranges. This value is attained for all values of coupling constants $|\Dd|$ and $\Jj$ for the $\kappa = 1/2$ while it is attained for specific values of the coupling constants in the other two cases. The maximal DC conductivity also matches the residual conductivity at $T=0$ for the $\kappa =1/2$ and $\kappa = 1$ cases, while it happens at a finite temperature for the $\kappa = 2$ case. We argued in the introduction as well as Section \ref{a note on scaling arguments} that scaling arguments naively leads to a prediction of conducting and insulating transport properties for the $\kappa =1/2$ and $\kappa = 1$ cases, respectively. However, the scaling arguments are not rigorous away from criticality for SYK-like systems but can only be taken as a guiding tool and not as a mathematical substitute for all the calculations that we have performed in this work. We did confirm the naive expectations of scaling arguments and using this justification, we argued the universality of residual conductivity for all coupling constants for these two cases in Section \ref{scaling arguments and universality}.

We have tried to keep our formalism quite general that is amenable to different setups where the action is written in terms of $G-\Sigma$ formalism. An interesting extension of our work can be to include higher order correlations of current that can possibly relate to the analytic calculation of the Fano factor, which is taken as another mesoscopic signature of the strange metal behavior \cite{Nikolaenko2023Nov, Chen2023Nov}. Having analytical calculations at our disposal, quantum quench dynamics in such SYK chains \cite{Jha2023} might become feasible, where relaxation rates and thermalization can be studied in the thermodynamic limit \cite{Larzul2022Jan, Bhattacharya2019Jul}. 

A natural question to ask is the relation between quantum chaotic bound and maximal DC conductivity. We found for the three cases considered in this work a universal DC conductivity that seems to be connected to the maximum quantum chaos (in the sense of saturating the MSS bound \cite{Maldacena2016Aug}) because we get maximal DC conductivity when $v$ takes its maximum value, which implies a maximal quantum chaotic bound. Further, the conductivities we found  in the strange metal cases were directly proportional to the Lyapunov exponents. As such, an intriguing aspect of our findings is that this maximum conductivity is mathematically directly related to the MSS bound on chaos. Such Planckian dynamics is also seen in high-$T_c$ materials \cite{Legros2019Feb}. Given this, one might be inclined to see this as a relation between the scattering rate from the Drude model and the scrambling rate from the out-of-time-order correlators (OTOCs). However, a more in-depth analysis is required before claiming this to be more than an intriguing coincidence. We leave these questions as future lines of inquiry.

In this study we have strictly focused on the large-$q$ limit. A natural question to ask is to what extent these results remain for finite $q$. This can be addressed by considering next leading order 1/q corrections. However, given the overlap of the large $q$ solutions and the low temperature finite $q$ conformal solutions, we do not expect much to change, as has been noted in various transport properties studied in the $\kappa = 1$ cases \cite{Zanoci2022Jun}. However a direct calculation remains as an outlook for the current work.

In extending our study to next nearest neighbor interactions we also do not expect much to change aside from the continuity equations. Likely it would yield some increase in the conductivity due to the increased ease of hopping. Attempting to extend the study to non-uniform hopping would unfortunately mean that we no longer have the exact Green's functions, since we would have different Green's functions corresponding to different sites. The same problem also occurs if one has a charge or temperature gradient. In such a case, one could actually still treat the problem analytically using perturbation theory for a small non-uniformity \cite{Zanoci2022Jun} or resort to numerical techniques to solve the Schwinger-Dyson equations.

The exact nature of possible quasiparticles in the insulator type system ($\kappa=2$ case) discussed in Sec. \ref{FLSection} and the possibility of a deeper holographic connection discussed in Sec. \ref{HolographicInsulatorSection} remain open questions. Future studies should investigate whether our system adheres to the Wiedemann-Franz law in the intermediate temperature regime, which would further elucidate the presence and nature of Fermi-liquid-like quasiparticles.

Additionally, one can study the cubic corrections to Fermi liquid behavior in the insulating case. The resistivity has a quadratic dependency on temperature around the minimum value, namely $\rho \sim (T - T_{\text{min}})^2$ where $T_{\text{min}}$ is the temperature value where the minimum resistivity is attained for the $\kappa =2$ insulating case (see Fig. \ref{fig:Res}). Then any cubic corrections would lead to deviation from a Fermi liquid quasiparticle picture, which becomes negligibly small as $T \to T_{\text{min}}$. However, the size of the cubic corrections will provide an estimate to the existence of quasiparticles. These aspects warrant further theoretical and numerical investigations to fully understand the crossover behaviors and the underlying mechanisms.


\section*{Acknowledgment}
\label{acknowledgment}
RJ, SK and JCL would like to thank Deutsche Forschungsgemeinschaft (DFG, German Research Foundation)) - 217133147/SFB 1073 (Project No. B03) for supporting this work. JCL also acknowledges the financial support of the Max Planck Society, the Munich Center for Quantum Science and Technology, and the hosting of the Ludwig-Maximilians University, Munich as well as the University of Göttingen.

\bibliography{conductivity.bib}

\newpage

\appendix

\onecolumngrid

\section{Forward and backward contour deformations}
\label{app. forward and backward contour deformation}
We have
\begin{subequations}
	\begin{align}
		S_{I}[\Gg] &=  +\nint[\Cc]{t_1} \nint[\Cc]{t_2}\frac{D_{\Cc}(t_1)\bar{D}_{\Cc}(t_2) }{2}  F(\Gg(t_1,t_2)\Gg(t_2,t_1))   \\
		&= +\nint[\Cc]{t_1} \nint[\Cc]{t_2}\frac{D_{\Cc}(t_1)\bar{D}_{\Cc}(t_2) }{2}  \left( X (t_1, t_2) +\sgn_\Cc(t_1 - t_2) \i Y(t_1, t_2) \right) \\
		&= \Re S_I + \i \Im S_I \qquad (X(t_1, t_2) = X(t_2, t_1) \text{ and } Y(t_1, t_2) = - Y(t_2, t_1))
	\end{align}
 \label{interacting action general}
\end{subequations}
where $D_\Cc(t) = D_-(t) + D_+(t)$ and $\bar{D}_\Cc(t) = \bar{D}_-(t) + \bar{D}_+(t)$.

We deform now both the forward and the backward contours as follows:
\begin{subequations}
\begin{align}
    D_-(t) &= D(t) + \Delta_-(t) = D(t)+ \sum_n \ll_n D(\tau_n) \delta(\tau_n - t), \\
    \bar{D}_-(t) &= D^*(t) + \bar{\Delta}_-(t) =  D^*(t) + \sum_n \llbar_n D^*(\tau_n) \delta(\tau_n - t) , \\
    D_+(t) &= D(t)- \Delta_+(t) = D(t) - \sum_n \et_n D(\tau_n) \delta(\tau_n - t), \\
    \bar{D}_+(t) &= D^*(t) - \bar{\Delta}_+(t)= D^*(t) - \sum_n \etbar_n D^*(\tau_n) \delta(\tau_n - t) 
    \end{align}
    \label{deformation contours back and forth 2}
\end{subequations}

The Keldysh contour along $\Cc_-$ runs from $t_0$ to $t_f$ while the contour $\Cc_+$ runs in reverse from $t_f$ to $t_0$ which we take into account with a minus sign as follows:
\begin{subequations}
	\begin{align}
		\nint[\Cc_-][]{t} (\cdots) &= \nint[t_0][t_f]{t} (\cdots)  \\
		\nint[\Cc_+][]{t} (\cdots) &= \nint[t_f][t_0]{t} (\cdots) = -\nint[t_0][t_f]{t} (\cdots) 
	\end{align}
\end{subequations}

We start from the following point for the real part of the interacting action:
\begin{equation}
    \begin{aligned}
    \Re S_I =& \nint[\Cc]{t_1} \nint[\Cc]{t_2}\frac{D_{\Cc}(t_1)\bar{D}_{\Cc}(t_2)}{2} X(t_1,t_2)\\
		=&  \nint[\Cc_+]{t_1} \nint[\Cc_-]{t_2}\frac{D_+(t_1)\bar{D}_-(t_2)}{2} X(t_1,t_2) + \nint[\Cc_-]{t_1} \nint[\Cc_+]{t_2}\frac{D_-(t_1)\bar{D}_+(t_2)}{2} X(t_1,t_2)\\
		&  \nint[\Cc_+]{t_1} \nint[\Cc_+]{t_2}\frac{D_+(t_1)\bar{D}_+(t_2)}{2} X(t_1,t_2) + \nint[\Cc_-]{t_1} \nint[\Cc_-]{t_2}\frac{D_-(t_1)\bar{D}_-(t_2)}{2} X(t_1,t_2) \\
		=&  \nint{t_1} \nint{t_2}\frac{-D_+(t_1) \bar{D}_-(t_2)-D_-(t_1)\bar{D}_+(t_2)+D_+(t_1)\bar{D}_+(t_2)+D_-(t_1)\bar{D}_-(t_2)}{2} X(t_1,t_2) \\
    =&\frac{1}{2} \nint{t_1}\nint{t_2} X(t_1, t_2) \left[ D_-(t_1) \bar{D}_-(t_2) + D_+(t_1) \bar{D}_+(t_2) - D_-(t_1) \bar{D}_+(t_2)- D_+(t_1) \bar{D}_-(t_2)\right]
    \end{aligned}
\end{equation}
where we plug in Eq. \eqref{deformation contours back and forth 2} to get
\begin{equation}
  \Re S_I =\Re S_I |_{\vec{\ll}, \vec{\et} = 0} + \frac{1}{2}  \nint{t_1}\nint{t_2} X(t_1, t_2) \left[\Delta_-(t_1) \bar{\Delta}_-(t_2) + \Delta_+(t_1) \bar{\Delta}_+(t_2) + \Delta_-(t_1) \bar{\Delta}_+(t_2) + \Delta_+(t_1) \bar{\Delta}_-(t_2) \right]         
\end{equation}
where $\vec{\ll} = \left( \{\ll_n \}, \{\llbar_m \} \right)$, $\vec{\et} = \left( \{\et_n \}, \{\etbar_m \} \right)$ and we plug the definitions of $\Delta_i$ and $\bar{\Delta}_i$ ($i = \{-, + \}$) from Eq. \eqref{deformation contours back and forth 2} to get
\begin{equation}
\begin{aligned}
    \Re S_I &= \Re S_I |_{\vec{\ll}, \vec{\et} = 0} + \frac{1}{2} \sum\limits_{n, m} D(\tau_n) D^*(\tau_m)  \left[ \ll_n \llbar_m + \et_n \etbar_m + \ll_n \etbar_m + \et_n \llbar_m  \right] X(\tau_n, \tau_m) \\
    &= \Re S_I |_{\vec{\ll}, \vec{\et} = 0} \\
    &+ \frac{1}{2} \sum\limits_{n> m} X(\tau_n, \tau_m)\left\{  D(\tau_n) D^*(\tau_m)  \left[ \ll_n \llbar_m + \et_n \etbar_m + \ll_n \etbar_m + \et_n \llbar_m  \right]  + D(\tau_m) D^*(\tau_n)  \left[ \ll_m \llbar_n + \et_m \etbar_n + \ll_m \etbar_n + \et_m \llbar_n  \right] \right\} \\
    &+\frac{1}{2} \sum\limits_{n} |D(\tau_n)|^2 \left[ |\ll_n|^2 + |\et_n|^2 +\left( \ll_n \etbar_n + \et_n \llbar_n \right)\right] X(\tau_n, \tau_n) 
    \end{aligned}
\end{equation}
which can be rearranged to finally get
\begin{equation}
    \boxed{\begin{aligned}
    \Re S_I = & \Re S_I |_{\vec{\ll}, \vec{\et} = 0} + \sum\limits_{n> m}  D(\tau_n) D^*(\tau_m)  \left[ \ll_n \llbar_m + \et_n \etbar_m + \ll_n \etbar_m + \et_n \llbar_m  \right] X(\tau_n, \tau_m) \\
    &+ \frac{1}{2} \sum\limits_{n} |D(\tau_n)|^2 \left[ |\ll_n|^2 + |\et_n|^2 +\left( \ll_n \etbar_n + \et_n \llbar_n \right)\right] X(\tau_n, \tau_n) 
    \end{aligned}}
    \label{real action total 2}
\end{equation}
or equivalently
\begin{equation}
\boxed{\begin{aligned}
\Re & S_I  =\Re S_I |_{\vec{\ll}, \vec{\et} = 0} \\
&+ \frac{1}{2} \sum\limits_{n= m} X(\tau_n, \tau_m)\left\{  D(\tau_n) D^*(\tau_m)  \left[ \ll_n \llbar_m + \et_n \etbar_m + \ll_n \etbar_m + \et_n \llbar_m  \right]  + D(\tau_m) D^*(\tau_n)  \left[ \ll_m \llbar_n + \et_m \etbar_n + \ll_m \etbar_n + \et_m \llbar_n  \right] \right\} \\
&+\frac{1}{2} \sum\limits_{n} |D(\tau_n)|^2 \left[ |\ll_n|^2 + |\et_n|^2 +\left( \ll_n \etbar_n + \et_n \llbar_n \right)\right] X(\tau_n, \tau_n) 
    \end{aligned}}
\end{equation}

Next we evaluate the imaginary part of the interacting action as follows:
\begin{equation}
    \begin{aligned}
        \Im S_I =& \frac{1}{2}\nint{t_1} \nint{t_2} \sgn(t_1 - t_2) Y(t_1,t_2) \left[  D_-(t_1) \bar{D}_-(t_2) -  \bar{D}_+(t_1) D_+(t_2)\right] \\
    &+ \frac{1}{2}\nint{t_1} \nint{t_2} Y(t_1,t_2) \left[  D_-(t_1) \bar{D}_+(t_2) -  D_+(t_1) \bar{D}_-(t_2)\right] \\
    =& \Im S_I |_{\vec{\ll}, \vec{\et} = 0} + \frac{1}{2}\nint{t_1}\nint{t_2} \sgn(t_1 - t_2) Y(t_1, t_2) \\
    &\left[ D(t_1) \bar{\Delta}_-(t_2) + D^*(t_1) \Delta_-(t_2) + D^*(t_1) \Delta_+(t_2) + D(t_1) \bar{\Delta}_+(t_2) + \Delta_-(t_1) \bar{\Delta}_-(t_2) - \Delta_+(t_1) \bar{\Delta}_+(t_2) \right] \\
    -& \frac{1}{2} \nint{t_1}\nint{t_2} Y(t_1, t_2) \left[ D(t_1) \bar{\Delta}_+(t_2) +D^*(t_1) \Delta_-(t_2) + D^*(t_1) \Delta_+(t_2) + D(t_1) \bar{\Delta}_-(t_2) + \Delta_-(t_1) \bar{\Delta}_+(t_2) - \Delta_+(t_1) \bar{\Delta}_-(t_2) \right] \\
    =& \Im S_I |_{\vec{\ll}, \vec{\et} = 0} - \frac{1}{2} \nint{t_1}\nint{t_2} Y(t_1, t_2) (1 - \sgn(t_1-t_2) \left[ D(t_1) \bar{\Delta}_-(t_2) + D^*(t_1) \Delta_-(t_2) + D^*(t_1) \Delta_+(t_2) + D(t_1) \bar{\Delta}_+(t_2)   \right] \\
    &+  \frac{1}{2} \nint{t_1} \nint{t_2} \sgn(t_1 - t_2) Y(t_1, t_2) \left[ \Delta_-(t_1) \bar{\Delta}_-(t_2) - \Delta_+(t_1) \bar{\Delta}_+(t_2)\right] \\
    &-  \frac{1}{2} \nint{t_1}\nint{t_2}Y(t_1, t_2) \left[\Delta_-(t_1) \bar{\Delta}_+(t_2) - \Delta_+(t_1) \bar{\Delta}_-(t_2) \right]
    \end{aligned}
\end{equation}
We evaluate each of the terms separately by plugging the definitions of $\Delta_j$ and $\bar{\Delta}_j$ where $j =\{-, +\}$ from Eq. \eqref{deformation contours back and forth 2} and the identity $1 - \sgn(t_1 - t_2) = 2 \Theta(t_2 -t_1)$ to get
\begin{equation}
    \begin{aligned}
        - \frac{1}{2} \nint{t_1}\nint{t_2} &Y(t_1, t_2) (1 - \sgn(t_1-t_2) )\left[ D(t_1) \bar{\Delta}_-(t_2) + D^*(t_1) \Delta_-(t_2) + D^*(t_1) \Delta_+(t_2) + D(t_1) \bar{\Delta}_+(t_2)   \right] \\
        &= -\sum_m \nint[t_0][\tau_m]{t_1}  Y(t_1, \tau_m) \left[ D(t_1) D^*(\tau_m) \llbar_m + D^*(t_1) D(\tau_m) \ll_m + D^*(t_1) D(\tau_m) \et_m + D(t_1) D^*(\tau_m) \etbar_m \right],
    \end{aligned}
\end{equation}
\begin{equation}
    \begin{aligned}
        + \frac{1}{2}\nint{t_1} \nint{t_2} &\sgn(t_1 - t_2) Y(t_1, t_2) \left[ \Delta_-(t_1) \bar{\Delta}_-(t_2) - \Delta_+(t_1) \bar{\Delta}_+(t_2)\right] \\
        =& \frac{1}{2}\sum_{n,m} \sgn(\tau_n - \tau_m) Y(\tau_n, \tau_m) D(\tau_n) D^*(\tau_m) \left[ \ll_n \llbar_m - \et_n \etbar_m \right] \\
        =& \frac{1}{2} \sum\limits_{n>m} \left[ Y(\tau_n, \tau_m) D(\tau_n) D^*(\tau_m) \left[ \ll_n \llbar_m - \et_n \etbar_m \right] - Y(\tau_m, \tau_n) D(\tau_m) D^*(\tau_n) \left[ \ll_m \llbar_n - \et_m \etbar_n \right] \right]\\
        =& \frac{1}{2} \sum\limits_{n>m} Y(\tau_n, \tau_m) \left[ D(\tau_n) D^*(\tau_m) \left[ \ll_n \llbar_m - \et_n \etbar_m \right] +  D(\tau_m) D^*(\tau_n) \left[ \ll_m \llbar_n - \et_m \etbar_n \right] \right] 
    \end{aligned}
\end{equation}
and
\begin{equation}
    \begin{aligned}
 - \frac{1}{2}\nint{t_1}\nint{t_2}&Y(t_1, t_2) \left[\Delta_-(t_1) \bar{\Delta}_+(t_2) - \Delta_+(t_1) \bar{\Delta}_-(t_2) \right]      \\
 =&-\frac{1}{2}\sum\limits_{n,m} Y(\tau_n, \tau_m) D(\tau_n) D^*(\tau_m) \left[ \ll_n \etbar_m - \et_n \llbar_m \right] \\
 =&-\frac{1}{2}\sum\limits_{n>m} \left[ Y(\tau_n, \tau_m) D(\tau_n) D^*(\tau_m) \left[ \ll_n \etbar_m - \et_n \llbar_m \right]+Y(\tau_m, \tau_n) D(\tau_m) D^*(\tau_n) \left[ \ll_m \etbar_n - \et_m \llbar_n \right] \right] \\
  =&-\frac{1}{2}\sum\limits_{n>m} Y(\tau_n, \tau_m) \left[ D(\tau_n) D^*(\tau_m) \left[ \ll_n \etbar_m - \et_n \llbar_m \right]- D(\tau_m) D^*(\tau_n) \left[ \ll_m \etbar_n - \et_m \llbar_n \right]\right]  
    \end{aligned}
\end{equation}

Therefore we have for the imaginary part of the interacting action
\begin{equation}
    \boxed{
\begin{aligned}
    \Im & S_I = \Im S_I |_{\vec{\ll}, \vec{\et} = 0}\\ 
    &+ \frac{1}{2}\sum\limits_{n>m} Y(\tau_n, \tau_m) \left\{ D(\tau_n) D^*(\tau_m) \left[ \ll_n \llbar_m - \et_n \etbar_m  -\ll_n \etbar_m + \et_n \llbar_m  \right] +  D(\tau_m) D^*(\tau_n) \left[ \ll_m \llbar_n - \et_m \etbar_n +\ll_m \etbar_n - \et_m \llbar_n  \right] \right\} \\
    &-\sum_m \nint[t_0][\tau_m]{t_1}  Y(t_1, \tau_m) \left[ D(t_1) D^*(\tau_m) \llbar_m + D^*(t_1) D(\tau_m) \ll_m + D^*(t_1) D(\tau_m) \et_m + D(t_1) D^*(\tau_m) \etbar_m \right]
\end{aligned}
    }
    \label{imaginary action total 2}
\end{equation}

Therefore the total interacting action is $S_I = \Re S_I + \i \Im S_I$ where $\Re S_I$ and $\Im S_I$ are given in eqs. \eqref{real action total 2} and \eqref{imaginary action total 2}, respective. We have $F(t_1, t_2) = X(t_1, t_2) + \i Y(t_1, t_2)$ and $F^*(t_1, t_2) = X(t_1, t_2) - \i Y(t_1, t_2)$ to get
\begin{equation}
    \begin{aligned}
 S_I &= S_I |_{\vec{\ll}, \vec{\et} = 0} \\
 &+ \frac{1}{2} \sum\limits_{n> m} X(\tau_n, \tau_m)\left\{  D(\tau_n) D^*(\tau_m)  \left[ \ll_n \llbar_m + \et_n \etbar_m + \ll_n \etbar_m + \et_n \llbar_m  \right]  + D(\tau_m) D^*(\tau_n)  \left[ \ll_m \llbar_n + \et_m \etbar_n + \ll_m \etbar_n + \et_m \llbar_n  \right] \right\} \\
 &+\frac{1}{2}\i \sum\limits_{n>m} Y(\tau_n, \tau_m) \left\{ D(\tau_n) D^*(\tau_m) \left[ \ll_n \llbar_m - \et_n \etbar_m  -\ll_n \etbar_m + \et_n \llbar_m  \right] +  D(\tau_m) D^*(\tau_n) \left[ \ll_m \llbar_n - \et_m \etbar_n +\ll_m \etbar_n - \et_m \llbar_n  \right] \right\} \\
&- \i \sum_m \nint[t_0][\tau_m]{t_1}  Y(t_1, \tau_m) \left[ D(t_1) D^*(\tau_m) \llbar_m + D^*(t_1) D(\tau_m) \ll_m + D^*(t_1) D(\tau_m) \et_m + D(t_1) D^*(\tau_m) \etbar_m \right]\\
&+ \frac{1}{2} \sum\limits_{n} |D(\tau_n)|^2 \left[ |\ll_n|^2 + |\et_n|^2 +\left( \ll_n \etbar_n + \et_n \llbar_n \right)\right] X(\tau_n, \tau_n) \\
&= S_I |_{\vec{\ll}, \vec{\et} = 0} + \frac{1}{2} \sum\limits_{n> m} D(\tau_n) D^*(\tau_m) \left( \ll_n \llbar_m + \et_n \llbar_m \right) F(\tau_n, \tau_m) + \frac{1}{2} \sum\limits_{n> m} D(\tau_n) D^*(\tau_m) \left( \et_n \etbar_m + \ll_n \etbar_m \right) F^*(\tau_n, \tau_m) \\
&+ \frac{1}{2} \sum\limits_{n> m} D(\tau_m) D^*(\tau_n) \left( \ll_m \llbar_n + \ll_m \etbar_n \right) F(\tau_n, \tau_m) + \frac{1}{2} \sum\limits_{n> m} D(\tau_m) D^*(\tau_n) \left(  \et_m \etbar_n + \et_m \llbar_n \right) F^*(\tau_n, \tau_m) \\
&- \i \sum_m \nint[t_0][\tau_m]{t_1}  Y(t_1, \tau_m) \left[ D(t_1) D^*(\tau_m) \llbar_m + D^*(t_1) D(\tau_m) \ll_m + D^*(t_1) D(\tau_m) \et_m + D(t_1) D^*(\tau_m) \etbar_m \right]\\
&+ \frac{1}{2} \sum\limits_{n} |D(\tau_n)|^2 \left[ |\ll_n|^2 + |\et_n|^2 +\left( \ll_n \etbar_n + \et_n \llbar_n \right)\right] X(\tau_n, \tau_n) 
    \end{aligned}
\end{equation}
So we have finally for the full interacting Hamiltonian
\begin{equation}
    \boxed{
    \begin{aligned}
      S_I =&   S_I |_{\vec{\ll}, \vec{\et} = 0} + \frac{1}{2} \sum\limits_{n> m} D(\tau_n) D^*(\tau_m) \left( \ll_n \llbar_m + \et_n \llbar_m \right) F(\tau_n, \tau_m) + \frac{1}{2} \sum\limits_{n> m} D(\tau_n) D^*(\tau_m) \left( \et_n \etbar_m + \ll_n \etbar_m \right) F^*(\tau_n, \tau_m) \\
&+ \frac{1}{2} \sum\limits_{n> m} D(\tau_m) D^*(\tau_n) \left( \ll_m \llbar_n + \ll_m \etbar_n \right) F(\tau_n, \tau_m) + \frac{1}{2} \sum\limits_{n> m} D(\tau_m) D^*(\tau_n) \left(  \et_m \etbar_n + \et_m \llbar_n \right) F^*(\tau_n, \tau_m) \\
&- \i \sum_m \nint[t_0][\tau_m]{t_1}  Y(t_1, \tau_m) \left[ D(t_1) D^*(\tau_m) \llbar_m + D^*(t_1) D(\tau_m) \ll_m + D^*(t_1) D(\tau_m) \et_m + D(t_1) D^*(\tau_m) \etbar_m  \right]\\
&+\frac{1}{2} \sum\limits_{n} |D(\tau_n)|^2 \left[ |\ll_n|^2 + |\et_n|^2 +\left( \ll_n \etbar_n + \et_n \llbar_n \right)\right] X(\tau_n, \tau_n) 
    \end{aligned}
    }
    \label{full interacting action in terms of lambda and eta}
\end{equation}

Now we impose \textit{equilibrium}, consider time-independent couplings and $t_0 \to -\infty$ to get
\begin{equation}
    \begin{aligned}
        \frac{S_I}{|D|^2} &= S_I |_{\vec{\ll}, \vec{\et} = 0} \\
        &+ \frac{1}{2} \sum\limits_{n> m} F(\tau_n - \tau_m) \left(\ll_n \llbar_m + \et_n \llbar_m + \ll_m \llbar_n + \ll_m \etbar_n \right) + \frac{1}{2} \sum\limits_{n> m} F^*(\tau_n - \tau_m) \left(\et_n \etbar_m + \ll_n \etbar_m +\et_m \etbar_n + \et_m \llbar_n \right) \\
        &- \i \sum_m \nint[-\infty][0]{t}  Y(t) \left[  \llbar_m + \ll_m + \et_m +  \etbar_m  \right] +\frac{1}{2} \sum\limits_{n} (\ll_n + \et_n)(\llbar_n + \etbar_n) X(0) \\
\Rightarrow  \frac{S_I}{|D|^2}        &= S_I |_{\vec{\ll}, \vec{\et} = 0} + \frac{1}{2} \sum\limits_{n>m} \left[ (\ll_n + \et_n) (\llbar_m F(\tau_n - \tau_m) + \etbar_m F^*(\tau_n - \tau_m) ) + (\llbar_n + \etbar_n)(\ll_m F(\tau_n - \tau_m)  +  \et_m F^*(\tau_n - \tau_m)) \right]  \\
        &- \i \sum_m \nint[-\infty][0]{t}  Y(t) \left[  \llbar_m + \ll_m + \et_m +  \etbar_m  \right] +\frac{1}{2} \sum\limits_{n} (\ll_n + \et_n)(\llbar_n + \etbar_n) X(0) 
    \end{aligned}
\end{equation}
Therefore result for the equilibrium case where we have considered time-independent couplings and $t_0 \to -\infty$ result is
\begin{equation}
    \boxed{
    \begin{aligned}
        \frac{S_I}{|D|^2}  =& S_I |_{\vec{\ll}, \vec{\et} = 0} + \frac{1}{2} \sum\limits_{n>m} \left[ (\ll_n + \et_n) (\llbar_m F(\tau_n - \tau_m) + \etbar_m F^*(\tau_n - \tau_m) ) + (\llbar_n + \etbar_n)(\ll_m F(\tau_n - \tau_m)  +  \et_m F^*(\tau_n - \tau_m)) \right]  \\
        &- \i \sum_m \left[  \llbar_m + \ll_m + \et_m +  \etbar_m  \right] \left( \nint[-\infty][0]{t}  Y(t) \right) +\frac{1}{2} \sum\limits_{n} (\ll_n + \et_n)(\llbar_n + \etbar_n) X(0)  \qquad (\text{Equilibrium})
    \end{aligned}
    }
    \label{equilibrium effective action in terms of lambda and eta 2}
\end{equation}

\twocolumngrid

\section{Initial condition for \texorpdfstring{$\sigma$}{}}
\label{app. initial condition for sigma}

    To investigate this, we consider the explicit commutation relation as given in Eq. \eqref{sigma in terms of X commutator} (where we use $\Theta(0^+) = 1$)
    \begin{equation}
        \sigma(0) = \i \ex{[\hat{X}(0),\hat{I}_x(0)]}= \i \frac{\ex{[\hat{X}(0),\hat{I}(0)]}}{L}
        \label{gx def}
    \end{equation}
    where the final equality is due to translation invariance. 
    We have defined the polarization operator $\hat{X}$ in Eq. \eqref{polarization operator definition} and local charge density in Eq. \eqref{local charge density definition} which we reproduce here for convenience
    \begin{equation}
        \hat{X} = \sum_{j=1}^{L} j N \hat{\Qq}_j,\quad \hat{\Qq}_i = \frac{1}{N}\sum_{\alpha=1}^N \left( c_{i,\alpha}^\dag c_{i,\alpha}-\frac{1}{2}\right)
    \end{equation}
 The total current operator is given in Eq. \eqref{currentOp} as
$\hat{I} =  \i r( \hat{\Hh}_{\rightarrow} - \hat{\Hh}_{\rightarrow}^\dag)/2$.
We further know from the general Galitskii-Migdal relations (Eqs. \eqref{Galit1} and \eqref{Galit2}) that
\begin{equation}
\begin{aligned}
    [N\hat{\Qq}_{i},\hat{\Hh}_{i\to i+1}] &= -r \hat{\Hh}_{i\to i+1}/2\\
    [N\hat{\Qq}_{i},\hat{\Hh}_{i-1\to i}] &= r \hat{\Hh}_{i-1\to i}/2
    \end{aligned}
\end{equation}
Therefore we have
\begin{align}
    [N\hat{\Qq}_{i},\hat{\Hh}_{\rightarrow}] 
    &= r[\hat{\Hh}_{i-1\to i}-\hat{\Hh}_{i\to i+1}]/2\\
\end{align}
where $\hat{\Hh}_{\rightarrow}$ is defined in Eq. \eqref{H rightarrow defined}. For the local current operator as defined in Eq. \eqref{local current def} reproduced here for convenience (recall that $r = \kappa q$ in our case)
\be
\hat{I}_{i+1} = \i r\frac{\hat{\Hh}_{i\to i+1}-\hat{\Hh}_{i\to i+1}^\dag}{2},
\ee
we have
\begin{align}
        [N\hat{\Qq}_{i},\hat{I}_i] &= \i r[N\hat{\Qq}_{i}, \hat{\Hh}_{i-1\to i}-\hat{\Hh}_{i-1\to i}^\dag]/2\\
        &= \i r[N\hat{\Qq}_{i}, \hat{\Hh}_{i-1\to i}]/2-\i r[N\hat{\Qq}_{i}, \hat{\Hh}_{i-1\to i}^\dagger]/2\\        
        &= \i \frac{r^2}{4} \hat{\Hh}_{i-1\to i}+\i \frac{r^2}{4} \hat{\Hh}_{i-1\to i}^\dagger\\
            &= \i \frac{r^2}{4}\hat{\Hh}^{i-1, i}
\end{align}
where $\hat{\Hh}^{i-1, i}$ is defined in Eq. \eqref{H i, i+1 defined}. Therefore, using these results, we finally get
\begin{equation}
            [\hat{X},\hat{I}]
            = \frac{r^2}{4}\i \hat{\Hh}_{\text{trans}}
\end{equation}
where $\hat{\Hh}_{\text{trans}}$ is defined in Eq. \eqref{H dot and H trans defined}. This provides us with the initial condition for $\sigma$ that is given in Eq. \eqref{initial condition for sigma equation}
        \begin{equation}
        \begin{aligned}
        \sigma(0) &= \i \frac{\ex{[\hat{X}(0),\hat{I}(0)]}}{L} =  -\frac{r^2}{4} \frac{\ex{ \hat{\Hh}_{\text{trans}}}}{L} \\
        &= 2\,  \Im \nint[-\infty][0]{t}\ex{\hat{I}(0)\hat{I}(t)}/L 
        \end{aligned}
        \label{InitialGx}
    \end{equation}
where we have used Eq. \eqref{Etrans} to evaluate 
\begin{equation}
    \ex{\hat{\Hh}_{\text{trans}}} = \ex{\hat{\Hh}_{\rightarrow}} + \ex{\hat{\Hh}_{\rightarrow}^\dagger} = -2 \Nn L |D|^2 \nint[-\infty][0]{t} Y(t) .
\end{equation}
However, using Eq. \eqref{current correlation in terms of F main text}, we have
\begin{equation}
    \ex{\hat{I}(0)\hat{I}(t)} = + \frac{r^2}{4} NL |D|^2 F(t)
\end{equation}
therefore we get
\begin{equation}
     \ex{\hat{\Hh}_{\text{trans}}} = - \frac{8}{r^2} \Im   \nint[-\infty][0]{t}\ex{\hat{I}(0)\hat{I}(t)}
\end{equation}
This is what we plug in the first line of Eq. \eqref{InitialGx} to get the initial condition for $\sigma$. We have derived from the first principles that also served as the consistency check for the initial condition that can be read-off directly from the dynamical conductivity formula in Eq. \eqref{dynamical conductivity formula}.

\section{Fluctuation-dissipation theorem}
\label{app. fluctuation dissipation theorem}

 Let us consider the susceptibility that we have defined below Eq. \eqref{equation 85}
 \be 
 \chi^R(t) \equiv \dot{\sigma}(t) = \Theta(t) \frac{2\, \Im\ex{\hat{I}(0)\hat{I}(t)}}{L} 
 \label{chiR def in appendix}
 \ee
 where we have used Eq. \eqref{sigma dot equation} for $\dot{\sigma}$. We also have the following identity: 
 \be
 \begin{aligned}
 \ex{\hat{I}(0) \hat{I}(t)} &= \text{Tr}\{\hat{I}(0)\hat{I}(t)\rho\} = \text{Tr}\{ (\rho \hat{I}(t) \hat{I}(0))^\dag\} \\
 &= \ex{\hat{I}(t)\hat{I}(0)}^* 
 \label{complex conjugate in appendix}
 \end{aligned}
 \ee
 This allows us to write $\chi^R(t)$ as
 \be
 \chi^R(t) = \Theta(t) \frac{2\, \Im\ex{\hat{I}(0)\hat{I}(t)}}{L} = - \Theta(t) \frac{2\, \Im\ex{\hat{I}(t)\hat{I}(0)}}{L} 
 \ee
In terms of the current-current correlation functions (Eq. \eqref{pi^R definition})
\be
\Pi^R(t) \equiv \Theta(t)\frac{\ex{\hat{I}(t)\hat{I}(0)}}{\i L}   = \Theta(t)\frac{\ex{\hat{I}(0)\hat{I}(-t)}}{\i L},
\label{piR def in appendix}
\ee
we have 
\be \chi^R(t) = - 2\, \Re\Pi^R(t). \label{chiR in temrs of PiR}\ee

Recall that the superscript $R$ denotes the retarded function and that's why such functions come with $\Theta$-functions in their definitions. The relation in Eq. \eqref{chiR in temrs of PiR} also holds true without the requirement of retarded function, namely
\be \chi(t) = - 2\, \Re\Pi(t). \label{chi in temrs of PiR}\ee

The fluctuation dissipation theorem is based the following identity:
\begin{widetext}
\be
\ex{\hat{I}(0) \hat{I}(t)}=\text{Tr }\{e^{-\beta \Hh}\hat{I}(0)\hat{I}(t)\}=\text{Tr }\{ \hat{I}(t) e^{-\beta \Hh}\hat{I}(0)\}=\text{Tr }\{e^{-\beta \Hh}\underbrace{e^{\beta \Hh}\hat{I}(t) e^{-\beta \Hh}}_{\hat{I}(t-\i\beta)}\hat{I}(0)\}=\ex{\hat{I}(t-\i\beta) \hat{I}(0)}
\ee
\end{widetext}
which means that we have 
\begin{widetext}
\be
\chi(t) =  -  \frac{2\, \Im\ex{\hat{I}(t)\hat{I}(0)}}{L} = - \frac{\ex{\hat{I}(t)\hat{I}(0)}}{\i L} + \frac{\ex{\hat{I}(0)\hat{I}(t)}}{\i L} = - \Pi(t) +\frac{\ex{\hat{I}(t-\i\beta) \hat{I}(0)}}{\i L}
\ee
\end{widetext}
Taking Fourier transform on both sides leads to 
\begin{widetext}
\begin{equation}
\begin{aligned}
    \nint[-\infty][\infty]{t} e^{\i \omega t}\chi(t) &= -\Ff[\Pi](\omega) +\nint[-\infty][\infty]{t} e^{\i \omega t} \frac{\ex{\hat{I}(t-\i\beta) \hat{I}(0)}}{\i L}= -\Ff[\Pi](\omega) +e^{-\beta \omega}\nint[-\infty][\infty]{t} e^{\i \omega (t-\i\beta)} \frac{\ex{\hat{I}(t-\i\beta) \hat{I}(0)}}{\i L}\\
    &= -\Ff[\Pi](\omega) +e^{-\beta \omega}\nint[-\infty][\infty]{\tau} e^{\i \omega \tau} \frac{\ex{\hat{I}(\tau) \hat{I}(0)}}{\i L}
    \end{aligned}
\end{equation}
\end{widetext}
which leads to the fluctuation-dissipation theorem 
 \be 
 -\frac{\Im\chi(\omega)}{\omega} =\frac{ [  1 - e^{-\beta \omega}]}{\omega}\Im \Ff[\Pi](\omega)  . \label{fd 1 in appendix} \ee

But we wish to express this relation in terms of the retarded functions, namely $\chi^R$ and $\Pi^R$. We start by realizing that $\chi(t)$ is a real function that satisfies $\chi(-t) = -\chi(t)$ using Eqs. \eqref{chiR def in appendix} and \eqref{complex conjugate in appendix}. Then we get (recall that $\Theta(t)$ in $\chi^R(t)$ will make the Fourier integral over time to start from $t=0$)
\begin{equation}
\begin{aligned}
    \Im \chi^R(\omega) &= \nint[0][\infty]{t} \sin(\omega t) \chi(t) = \frac{1}{2}\nint[-\infty][\infty]{t} \sin(\omega t) \chi(t) \\
    &= \frac{1}{2} \Im \chi(\omega)
    \end{aligned}
    \label{relation between chi and chiR}
\end{equation}

Next, we use the definition of $\Pi(t)$ (same as $\Pi^R(t)$ but without the $\Theta$-function in Eq. \eqref{piR def in appendix}) in conjunction with Eq. \eqref{complex conjugate in appendix} to get the following relation:
\begin{equation}
     \Pi(t)^\star = - \Pi(-t)
\end{equation}
which leads to the following two relations:
\begin{subequations}
    \begin{align}
        \Re \Pi(t) &= \frac{\Pi(t) + \Pi(t)^\star}{2} =  \frac{\Pi(t) - \Pi(-t)}{2} \\
        \Im \Pi(t) &= \frac{\Pi(t) - \Pi(t)^\star}{2 \i} = \frac{\Pi(t) + \Pi(-t)}{2 \i}.
    \end{align}
\end{subequations}
Therefore we have (recall that $\Theta(t)$ in $\Pi^R(t)$ will make the Fourier integral over time to start from $t=0$)
\begin{equation}
    \begin{aligned}
        \Im \Ff[\Pi^R] (\omega) =& \Im \nint[0][\infty]{t} e^{\i \omega t} \Pi(t) \\
        =& \nint[0][\infty]{t} \cos(\omega t) \frac{\Pi(t) + \Pi(-t)}{2 \i} \\
        &+ \nint[0][\infty]{t} \sin(\omega t) \frac{\Pi(t) - \Pi(-t)}{2} \\
        =& \frac{1}{2} \nint[-\infty][\infty]{t} \cos(\omega t) \Pi(t) \\
        &+ \frac{1}{2}\nint[-\infty][\infty]{t} \sin(\omega t) \Pi(t)
    \end{aligned}
\end{equation}
while we have for the Fourier transform of the imaginary part of $\Pi(t)$ as
\begin{equation}
    \begin{aligned}
        \Im \Ff[\Pi](\omega) =& \nint[-\infty][\infty]{t} \cos(\omega t) \frac{\Pi(t) + \Pi(-t)}{2 \i} \\
        &+ \nint[-\infty][\infty]{t} \sin(\omega t) \frac{\Pi(t) - \Pi(-t)}{2} \\
        =& \nint[-\infty][\infty]{t} \cos(\omega t) \Pi(t) + \nint[-\infty][\infty]{t} \sin(\omega t) \Pi(t)
    \end{aligned}
\end{equation}
Thus we have
\begin{equation}
    \Im \Ff[\Pi^R](\omega) = \frac{1}{2} \Ff[\Pi](\omega)
    \label{relation between pi and piR}
\end{equation}

So we finally divide Eq. \eqref{fd 1 in appendix} by a factor of $2$ and use the relations in Eqs. \eqref{relation between chi and chiR} and \eqref{relation between pi and piR} to get the fluctuation-dissipation theorem 
\begin{equation}
     -\frac{\Im\chi^R(\omega)}{\omega} =\frac{ [  1 - e^{-\beta \omega}]}{\omega}\Im \Ff[\Pi^R](\omega)  . \label{fd 2 in appendix} 
\end{equation}
which is the same as used in the main text (Eq. \eqref{f-d equation}).

\section{Evaluating current-current correlations}
\label{app. current-current correlation calculation}

We start with Eq. \eqref{current correlation in terms of derivatives} which we reproduce here for convenience
\begin{widetext}
    \begin{equation}
\ex{\hat{I}(0)\hat{I}(t)} = - \frac{r^2}{4} ( \underbrace{\p_{\ll_1}\p_{\bar{\ll}_2}}_{\circled{A}}+ \underbrace{\p_{\bar{\ll}_1}\p_{\ll_2}}_{\circled{B}}  - \underbrace{\p_{\bar{\ll}_1}\p_{\bar{\ll}_2}}_{\circled{C}} - \underbrace{\p_{\ll_1}\p_{\ll_2}}_{\circled{D}})\Zz|_{\vec{\ll} = 0}
\label{current-current correltation term by term}
\end{equation}
\end{widetext}

where we only consider the forward deformation of the Keldysh contour, as explained in the main text (Section \ref{current correlations section}). We re-write the forward deformed Keldysh action (the interacting part since the free part does not have any deformation) as
\onecolumngrid

\begin{equation}        
\frac{S_I}{|D|^2}  = S_I |_{\vec{\ll}= 0} + \frac{1}{2} \left[ \ll_2 \llbar_1 F(t_2 - t_1)  + \llbar_2 \ll_1 F(t_2 - t_1)   \right]  - \i \sum\limits_{m=1}^2 \left[  \llbar_m + \ll_m  \right] \left( \nint[-\infty][0]{t}  Y(t) \right) +\frac{1}{2} \left( \ll_1  \llbar_1 + \ll_2  \llbar_2  \right) X(0)
    \label{forward contour deformation in terms of labels 1 and 2}
\end{equation}
where we chose to focus and identify $\ll_1$ with $t_1 = 0$ and $\ll_2$ with $t_2 = t>0$ (see the paragraph below Eq. \eqref{Etrans} where we have cautioned about the time-ordering while calculating expectation values of type $\ex{\hat{A}(t_1) \hat{B}(t_2)}$ where $t_2>t_1$ for any operators $\hat{A}$ and $\hat{B}$).

We start by evaluating various terms in Eq. \eqref{current-current correltation term by term} as follows:
\begin{equation}
	\begin{aligned}
		\circled{A} = \p_{\ll_1} \p_{\bar{\ll}_2} \Zz|_{\vec{\ll}=0} &= \p_{\ll_1}\p_{\bar{\ll}_2} \int \mathcal{D}\Gg \mathcal{D}\Sigma e^{-\Nn L S(\ll)[\Gg, \Sigma]}\vert_{\vec{\ll} = 0} \\
		&=- NL  [ \p_{\ll_1} \p_{\bar{\ll}_2} S_I(\ll) ] |_{\vec{\ll}=0} + (NL)^2  \underbrace{[ \p_{\ll_1}S_I(\ll) ]|_{\vec{\ll}=0}  [  \p_{\bar{\ll}_2}S_I(\ll)]|_{\vec{\ll}=0}}_{\circled{A.1}} ,
	\end{aligned}
\label{mixed derivatives 1}
\end{equation} 
\begin{equation}
	\begin{aligned}
		\circled{B} = \p_{\bar{\ll}_1} \p_{\ll_2} \Zz|_{\vec{\ll}=0} &= \p_{\bar{\ll}_1}\p_{\ll_2} \int \mathcal{D}\Gg \mathcal{D}\Sigma e^{-\Nn L S(\ll)[\Gg, \Sigma]}\vert_{\vec{\ll}= 0} \\
		&=- NL [ \p_{\bar{\ll}_1} \p_{\ll_2} S_I(\ll) ] |_{\vec{\ll}=0} + (NL)^2 \underbrace{[ \p_{\bar{\ll}_1}S_I(\ll) ]|_{\vec{\ll}=0} [  \p_{\ll_2}S_I(\ll)] |_{\vec{\ll}=0}}_{=\circled{B.1}}
	\end{aligned}
\label{mixed derivatives 2}
\end{equation}
\begin{subequations}
	\begin{align}
\circled{C} = \p_{\ll_1}\p_{\ll_2} \Zz\vert_{\vec{\ll} = 0}
&= -\Nn L  [\underbrace{\p_{\ll_1}\p_{\ll_2} S_I(\ll)}_{=0}]\vert_{\vec{\ll} = 0} + (\Nn L)^2 \underbrace{ [\p_{\ll_1} S_I(\ll)]|_{\vec{\ll}=0}[\p_{\ll_2} S_I(\ll)]\vert_{\vec{\ll} = 0}}_{\circled{C.1}}, \quad \text{and}\\
\circled{D} = \p_{\bar{\ll}_1}\p_{\bar{\ll}_2} \Zz\vert_{\vec{\ll} = 0}
&= -\Nn L  [\underbrace{\p_{\bar{\ll}_1}\p_{\bar{\ll}_2} S_I(\ll)}_{=0}]\vert_{\vec{\ll} = 0} + (\Nn L)^2  \underbrace{[\p_{\bar{\ll}_1} S_I(\ll)]|_{\vec{\ll}=0}[\p_{\bar{\ll}_2} S_I(\ll)]\vert_{\vec{\ll} = 0}}_{\circled{D.1}}
	\end{align}
\end{subequations}
where we used the fact that un-deformed Keldysh partition function is unity (Eq. \eqref{keldysh partition function equals unity}). To evaluate $\circled{A.1} - \circled{D.1}$, we first calculate the following for particular values of $n$ and $m$:
\begin{equation}
		\p_{\ll_n} S(\ll)|_{\vec{\ll}  = 0} = - \i |D|^2 \int_{-\infty}^0 dt  Y(t), \quad 	\p_{\bar{\ll}_m} S(\ll)|_{\vec{\ll}  = 0} = - \i |D|^2 \int_{-\infty}^0 dt Y(t)
		\label{general mixed derivatives}
\end{equation}
where, for example, if we wish to calculate $\left. \left[ \p_{\ll_1}S(\ll) \right]\right|_{\vec{\ll}=0} \left. \left[  \p_{\bar{\ll}_2}S(\ll)\right] \right|_{\vec{\ll}=0}$ in $\circled{A.1}$ in Eq. \eqref{mixed derivatives 1}, we substitute $n=1, m=2$ in Eq. \eqref{general mixed derivatives} while to evaluate $\left. \left[ \p_{\bar{\ll}_1}S(\ll) \right]\right|_{\vec{\ll}=0} \left. \left[  \p_{\ll_2}S(\ll)\right] \right|_{\vec{\ll}=0}$ in $\circled{B.1}$ in Eq. \eqref{mixed derivatives 2}, we substitute $n=2, m=1$ in Eq. \eqref{general mixed derivatives}. Therefore, we find that 
\begin{equation}
    \circled{A.1} = \circled{B.1} = \circled{C.1} = \circled{D.1} = -  (NL)^2 |D|^4 \left[ \int_{-\infty}^0 dt Y(t) \right]^2
    \label{A.1 to D.1}
\end{equation}
\twocolumngrid
Plugging Eqs. \eqref{mixed derivatives 1} - \eqref{A.1 to D.1} in Eq. \eqref{current-current correltation term by term}, we see that terms that are of order $(NL)^2$ exactly get cancelled out, and we are left with the following simplified result for the current-current correlation:
\begin{equation}
\ex{\hat{I}(0)\hat{I}(t)} = \frac{r^2}{4} NL (  \left.  \p_{\ll_1} \p_{\bar{\ll}_2} S_I(\ll)  \right|_{\vec{\ll}=0} +  \left. \p_{\bar{\ll}_1} \p_{\ll_2} S_I(\ll) \right|_{\vec{\ll}=0} )
\end{equation}
Using Eq. \eqref{forward contour deformation in terms of labels 1 and 2}, we further evaluate 
\begin{subequations}
	\begin{align}
 \left. \left[ \p_{\ll_1} \p_{\bar{\ll}_2} S(\ll) \right] \right|_{\vec{\ll}=0}&= + \frac{|D|^2}{2} F(t),\\
  \left. \left[ \p_{\bar{\ll}_1} \p_{\ll_2} S(\ll) \right] \right|_{\vec{\ll}=0}  &= +\frac{|D|^2}{2} F(t)
	\end{align}
 \end{subequations}
to finally get for current-current correlation
\begin{equation}
	\boxed{ \ex{\hat{I}(0)\hat{I}(t)} = + \frac{r^2}{4} NL |D|^2 F(t)}
	\label{current correlation in terms of F}
\end{equation}
where $r=\kappa q$ in our case for which we have already evaluated $F(t)$ in Eq. \eqref{expression for F} which we reproduce here for convenience
\begin{equation}
    F(t) = \frac{1}{\kappa q^2 } e^{- 2\kappa q \Qq^2} e^{\kappa g^+(t)} = X(t) + \i Y(t)
\end{equation}
Therefore, we specialize Eq. \eqref{current correlation in terms of F} to our case by plugging this expression for $F(t)$ to get
\begin{equation}
\ex{\hat{I}(0)\hat{I}(t)} = +\frac{\kappa}{4} NL\underbrace{ |D|^2 e^{-2 \kappa q \Qq^2} }_{\equiv |\Dd|^2/2 = \Oo(q^0) 	}e^{\kappa g^+(t)} 
\end{equation}
where charge density scales as $\Qq = \Oo(1/\sqrt{q})$. Here, we have defined
\begin{equation}
 |\Dd|^2 \equiv   2\, |D|^2 e^{-2 \kappa q \Qq^2}
\end{equation}
in the same spirit as we defined $\Jj_q$ and $\Kk_{\kappa q}$ in Eq. \eqref{curly J definition} where $\Dd = \Oo(q^0)$ in the large-$q$ limit. Recall the mapping between the uniformly coupled chain at equilibrium in Eq. \eqref{model equation} to the dot in Eq. \eqref{model equivalent equation}, namely $J^2 \to J_q^2$ and $2 |D|^2 \to K_{\kappa q}^2$ (see below Eq. \eqref{model equivalent equation}). In terms of $\{\Jj, \Dd\} $ for the chain, the mapping becomes $\{ \Jj^2 \to \Jj_q^2, |\Dd|^2 \to \Kk_{\kappa q}^2 \}$. So we finally get for our case
\begin{equation}
    \boxed{\ex{\hat{I}(0)\hat{I}(t)}_{\kappa} = +\frac{\kappa}{8} NL |\Dd|^2e^{\kappa g^+(t)} }
    \label{current correlation for our case}
\end{equation}
where we have introduced a label $\kappa$ in the subscript to denote the three cases we are interested in. We know the solutions for $g^+(t)$ for $\kappa = \{1/2, 1, 2\} $ as given in Section \ref{analytical solution subsection}. Using the property $\ex{\hat{I}(t) \hat{I}(0)}^* = \ex{\hat{I}(0)\hat{I}(t)}$ and $\left(g^+(t)\right)^* = g^+(-t)$ (see below Eq. \eqref{g plus minus def}), we can also get the current-current correlation in reversed time-ordering as
\begin{equation}
    \ex{\hat{I}(t)\hat{I}(0)}_{\kappa} = +\frac{\kappa}{8} NL |\Dd|^2e^{\kappa g^+(-t)} 
    \label{current correlation for our case for main integral}
\end{equation}

\end{document}